\documentclass{article}
\RequirePackage[OT1]{fontenc}
\usepackage{amsthm,amsmath}
\usepackage[authoryear]{natbib}

\usepackage{graphicx,setspace,amsfonts,subcaption,bbm}
\usepackage{rotating,tikz,mathtools,url,lscape,longtable,soul}

\RequirePackage[colorlinks,citecolor=blue,urlcolor=blue]{hyperref}

\def\bnu{\mbox{\boldmath $\nu$}}

\def\bgamma{\mbox{\boldmath $\gamma$}}

\def\btheta{\mbox{\boldmath $\theta$}}

\def\bLambda{\mbox{\boldmath $\Lambda$}}

\def\bbeta{\mbox{\boldmath $\beeta$}}

\def\bzeta{\mbox{\boldmath $\zeta$}}
\def\beeta{\mbox{\boldmath $\eta$}}
\def\bSigma{\mbox{\boldmath $\Sigma$}}

\def\bSigma{\mbox{\boldmath $\Sigma$}}

\def\bI{\mbox{\boldmath $I$}}

\def\bgamma{\mbox{\boldmath $\gamma$}}

\def\bDelta{\mbox{\boldmath $\Delta$}}

\def\bZ{\mathbf{Z}}

\def\bbb{\mathbf{b}}

\def\bZ{\mathbf{Z}}

\def\bx{\mathbf{x}}
\def\bX{\mathbf{X}}

\usepackage[authoryear]{natbib}
\usepackage[affil-it]{authblk}

\makeatletter
\renewcommand\AB@affilsepx{\quad \protect\Affilfont}
\makeatother

\begin{document}
\title{\textbf{The coordinate-based meta-analysis of neuroimaging data}}
\date{\today}
\author[1]{Pantelis Samartsidis}
\author[2]{Silvia Montagna}
\author[3]{Timothy D.\ Johnson}
\author[4]{Thomas E.\ Nichols}
\affil[1]{MRC Biostatistics Unit, University of Cambridge}
\affil[2]{School of Mathematics, Statistics \& Actuarial Science, University of Kent}
\affil[3]{Department of Biostatistics, University of Michigan}
\affil[4]{Oxford Big Data Institute, Li Ka Shing Centre for Health Information and Discovery, Nuffield Department of Population Health, and Wellcome Centre for Integrative Neuroimaging, FMRIB, Nuffield Department of Clinical Neurosciences, University of Oxford}

\renewcommand\Authands{ and }
\maketitle

\begin{abstract}
Neuroimaging meta-analysis is an area of growing interest in statistics. The special characteristics of neuroimaging data render classical meta-analysis methods inapplicable and therefore new methods have been developed. We review existing methodologies, explaining the benefits and drawbacks of each. A demonstration on a real dataset of emotion studies is included. We discuss some still-open problems in the field to highlight the need for future research.
\end{abstract}

\section{Introduction}
\label{sec:introduction}
{\em Functional magnetic resonance imaging} (fMRI) has experienced a rapid growth over the past two decades and has lead to significant advances in our understanding of the human brain, including the differences in brain function between maternal and romantic love \citep{Bartels2004}, the effect of alcohol while peforming simulated driving \citep{Calhoun2012}, or the effect of doing nothing at all \citep{Cole2010}. 
The availability of MRI scanners, inexpensive computational resources and accessible analysis software has made fMRI an ubiquitous tool in psychology, neurology and psychiatry, in addition to new areas like neuromarketing and neuroeconomics.

Nevertheless, there are a variety of factors that limit the interpretability of fMRI results. 
The principal limitation is the small sample sizes typically used, leading to individual studies that suffer from low power and hence low reproducibility \citep{Button2013}. 
Some other concerns include high prevalence of false positives \citep{Wager2007}, poor reproducibility \citep{Raemaekers2007} and considerable heterogeneity in the analysis pipeline \citep{Carp2012}.
As a result, it is unsurprising that the validity of fMRI is being challenged in both the scientific \citep{Vul2009c} and popular \citep{Shermer2008} literature  (see \citep{Farah2014} for an even-handed review). 

Meta-analysis provides a way to address all of these limitations. 
Meta-analysis is the process of combining the results of independently conducted studies to increase power and obtain more reproducible findings than the original studies \citep{Hedges1985}. 
Meta-analysis of functional neuroimaging data is an active field of research, whose growth is facilitated by the constantly increasing body of literature in fMRI along with the limitations of single experiments. 
The goal of this paper is to review recent advances in fMRI meta-analysis, evaluate existing methods, and highlight the open problems that need to be investigated. We consider a meta-analytic dataset analysed in a number of ways and make the data freely available to aid other researchers trying a hand at this intriguing area.

The remainder of this manuscript is organised as follows. In Section \ref{sec:neuroback} we provide some background on neuroimaging and explain the special characteristics of neuroimaging studies that make meta-analysis in this setting challenging. 
In Section \ref{sec:cbma} we give a detailed description of the currently most popular methods for fMRI meta-analysis. In Section \ref{sec:evaluations} we proceed to an evaluation of existing methods, which is motivated by application to a real dataset of emotion studies. 
Finally, we discuss some possible directions for research in Section \ref{sec:challenges}. 
\section{Neuroimaging background}\label{sec:neuroback}
What follows is a very brief review of fMRI and the practical steps involved in a fMRI study.  
For a more detailed introduction, see \citet{Lindquist2008} for review of fMRI for statisticians, or \citet{Kim2012} for a detailed, technical review of the meaning of the fMRI signal;  \citet{Huettel2009} provide an accessible textbook treatment, while \citet{Poldrack2011} give a practical, data-analysis-oriented perspective.

The objective of a single fMRI study is to identify the neural correlates of a physical, mental or perceptual process. 
When neurons in a region of the brain increase their firing rate, there is an increased demand for oxygen which is met by a localised increase in blood flow. 
The magnetic resonance signature, or susceptibility, of oxygenated and de-oxygenated blood differs, and thus a MRI scanner can capture changes in local oxygenation. 
This mechanism is known as the {\em blood oxygenation level-dependent} (BOLD) effect.

During an fMRI acquisition, participants lie flat in the scanner and are asked to perform a series of tasks, such as viewing images or reading texts, while the MRI scanner measures the BOLD signal.
For each participant, the data takes the form of a time series of images, 3D snapshots of signal measurements all over the brain.  
The typical acquisition lasts 6-12 minutes, with data collected every 2 seconds, producing data on a grid with 2mm $\times$ 2mm spacing in-plane and 2mm-4mm  slices, producing anywhere from 40,000 to over 100,000 voxels (volume elements) in the brain. 
Note that this is quite coarse spatial resolution, and separate, fine-resolution images (e.g. 1mm $\times$ 1mm $\times$ 1mm) are also taken to depict individual's anatomy.

\begin{figure}[h]
\centerline{\includegraphics[scale=0.26]{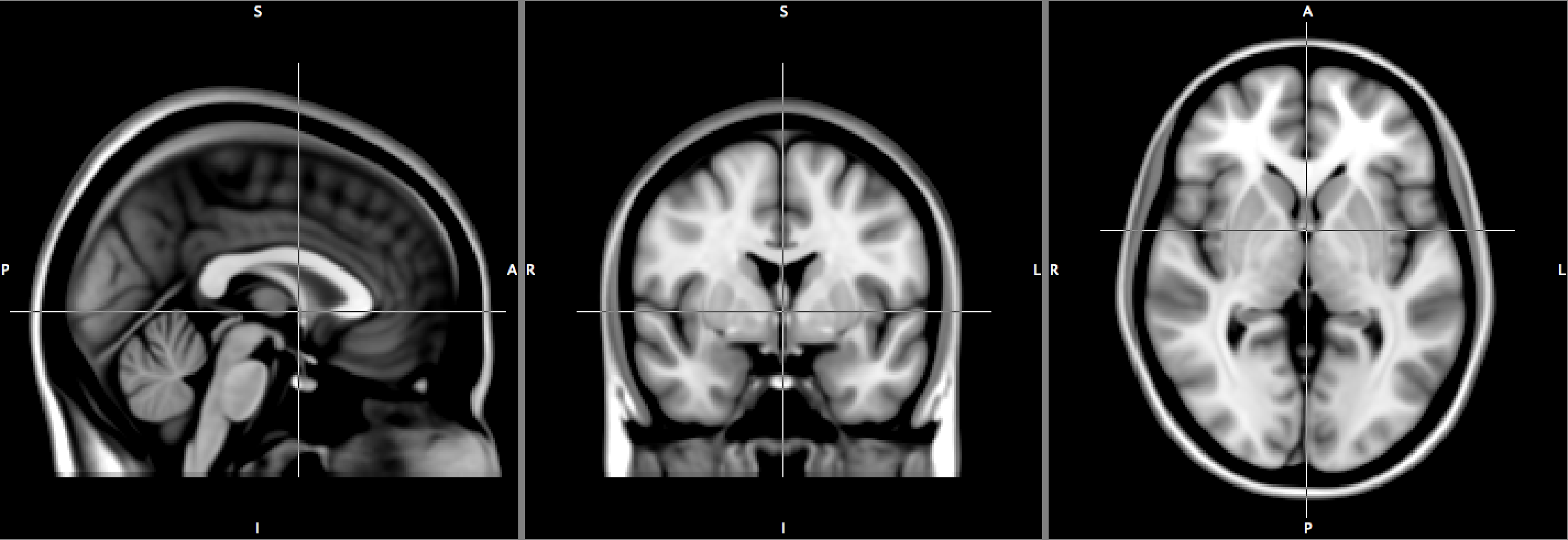}}
\caption{An average brain in MNI space. Note the directional labels at the edge of each panel: P for Posterior, A for Anterior, S for Superior, I for Inferior, L for Left and R for Right. The origin approximately corresponds to an anatomical structure known as the anterior commissure.}
\label{origin}
\end{figure}

Before the raw data can be analysed, a series of preprocessing steps needs to be undertaken. These include motion correction, which accounts for movements during the acquisition, and spatial smoothing which increases the signal-to-noise ratio.  
To make data comparable across subjects, a crucial step is ``spatial normalisation'', the process of warping all subjects to a standard brain template, or brain atlas.  
There are different atlases available, but essentially all authors use either the Talairach atlas \citep{Talairach1988} or the MNI atlas (see Fig.\ \ref{origin}).

After spatial normalisation, all subjects' data exist in a common space. 
Specifically, we can assume that a given voxel corresponds to (roughly) the same region in all subjects' brains. Statistical analysis then proceeds in a mass-univariate approach, fitting a model at each voxel independently of every other voxel. 
For every subject, time series regression models are fitted in each voxel, where the regression coefficients represent the effect of the different tasks. Task effects or comparisons can be made with the proper ``contrasts'', meaning the estimated linear combination of parameter estimates that relates to the effect of interest. These subject-specific contrasts are subsequently analysed in a ``second level'' population model. 
The result is a 3D image of $T$ statistics, one for each voxel in the brain, measuring the evidence against the null hypothesis of no effect.  
The $T$ images are assessed either voxel-by-voxel, or by assessing the size of connected components, or {\em clusters}, after thresholding the $T$ image at an arbitrary threshold.   
See \citet{Friston2002}, \citet{Mumford2006} and \citet{Mumford2009} for a detailed review of different approaches for the statistical analysis of fMRI data.

An essential issue in the statistical analysis of fMRI data is {\em multiple testing}. 
A $T$ statistic image can have 100,000 or more voxels in the brain, requiring 100,000 simultaneous tests for every contrast of interest. 
Under a global null hypothesis of no effect in any voxel, we therefore expect around 5,000 false positives using the classic significance level of $\alpha=0.05$.  
In the early history of fMRI (roughly 1992-2002), arbitrary rule-of-thumb thresholding procedures were common, like a combination of an uncorrected voxel-wise $\alpha=0.001$ and cluster size threshold $k\geq10$ (only clusters of size 10 voxels or more).  
Thresholding methods that controlled the {\em familywise error} (FWE), the chance of one or more false positives, later became widespread using either random field theory or permutation testing (see \citet{Hayasaka2003} for a review of FWE methods in neuroimaging). 
More recently, the {\em false discovery rate} (FDR) was introduced for the thresholding of $T$ images \citep{Genovese2002}. The FDR is the expected proportion of false positives among positive findings \citep{Benjamini1995}.

In any discipline of science, discussion of findings typically includes the point estimates for an effect, the associated standard errors, and the $p$-values. 
In neuroimaging, each of these quantities is 3D image, and sharing such large data files was considered impractical 20 years ago when fMRI was first developed. 
Yet even today there is general resistance towards sharing the full images. 
Instead, authors routinely report the $x,y,z$ atlas coordinates of activation peaks.  
Going forward we will call these coordinates the {\em foci} (singular {\em focus}).
In other words, the results of an fMRI study are summarised in a list of foci. 
Based on author preference and software defaults, foci can either be {\em singletons} that is one focus per significant region or {\em multiple} that is two or more per significant region.  

\subsection{Limitations of individual studies and meta-analysis}
\label{subsec:limitations}

There are three aspects of fMRI experiments that challenge the utility of individual studies. 
Firstly, fMRI studies suffer from low power. 
The typical sample size of an fMRI study is small, and the majority of experiments involves far less than 20 participants \citep{Carp2012}.  
While power depends on the (unknown) true effect size, at least one empirical study supported the notion that fMRI $n$'s are too small. 
By sub-sampling from a large sample ($n=150$), \citet{Thirion2007} found that analyses with 20 or fewer subjects were poor approximations of the full 150-subject result. 
Further, Type I error rates are likely to be high, especially for older papers that did not use  inference procedures corrected for multiple testing. 
Using a survey of publications' thresholding methods, \citet{Wager2007} estimated that 17\% of all reported foci are false positives. 
Finally, neuroimaging studies suffer from low test-retest reliability.  
For example, when scanning a group of subjects twice, once and then 7 days later, \citet{Raemaekers2007} found intra-class correlations for BOLD fMRI activations ranged from 0 to 0.88. 

Apart from these inherent limitations, the way fMRI studies are carried out also exhibits great heterogeneity. 
Each step of a neuroimaging study can be implemented in various ways and there is no standard way to present a stimulus, preprocess the data or construct the linear model for the BOLD response. 
As a result, there is only a partial agreement in how experiments are conducted. 
For example, in an analysis of 241 fMRI studies \citet{Carp2012} observed 223 different analytical strategies. 
Results heavily depend on the type of analysis employed \citep{Button2013}, thus it is not uncommon to observe discrepancies in the outcomes of studies that investigate the same scientific question. 
Consequently, it is exceptionally hard to yield a conclusion. 
All reasons combined support the use of meta-analysis to account for these problems and draw more reliable inferences.

A well performed meta-analysis can tackle the aforementioned issues by modelling the observed heterogeneity between studies, combining the available information to increase power and ultimately separating the consistent findings from those that happened by chance. 
There exist two broad approaches for meta-analysis of neuroimaging studies: {\em image-based meta-analysis} (IBMA), if the full $T$ statistic images are available, and {\em coordinate-based meta-analysis} (CBMA) if only foci are reported. 
IBMA proceeds by means of some common meta-analytic tools applied to each voxel of the images along with either FWER or FDR corrections for multiple testing. See \citet{Hartung2008} for an overview of conventional meta-analysis, and \citet{Lazar2002} for a review of IBMA methods. 

Given the richness of brain imaging data, with gigabytes collected on each subject, it is lamentable that all this data gets distilled to a list of $xyz$ coordinates (foci).
In a comparative study, \citet{Salimi2009} demonstrated the benefits of using IBMA over CBMA, and observed a dramatic loss of information when only foci were used relative to utilising full image data. 
Their results highlight the need for sharing full image data from single experiments (namely, effect magnitude estimates and standard errors, as well as key analysis details), and have encouraged the development of online repositories such as \textit{OpenfMRI}\footnote{\url{https://openfmri.org} (RRID:SCR\_005031)} \citep{Poldrack2013} and \textit{NeuroVault}\footnote{\url{http://neurovault.org} (RRID:SCR\_003806)} \citep{Gorgo2015} for sharing entire statistical maps. 
But sharing of full data is still rare, the use of such tools is not yet a standard practice and it is still not possible to access image data results for the majority of published research \citep{Poline2012}. 
On the contrary, coordinate data are easily available trough large databases such as \textit{BrainMap}\footnote{\url{http://www.brainmap.org} (RRID:SCR\_003069)} \citep{Laird2005a} and \textit{NeuroSynth}\footnote{\url{http://www.neurosynth.org} (RRID:SCR\_006798)} \citep{Yarkoni2011}. These repositories include on-line tools for quick data extraction and on-line implementation of CBMA meta-analyses.
As a result, CBMA still constitutes the main approach for the meta-analysis of fMRI data, and we will focus solely on a review of CBMA methods hereafter.

\section{CBMA methods} \label{sec:cbma}
The limitations of single experiments (see Section \ref{subsec:limitations} for a discussion), along with the historical lack of data sharing, quickly presented researchers in the field of fMRI with a challenge. 
The standard meta-analytic tools used in other fields (see for example \citet{Hartung2008} for a fairly recent review) could not be applied to the coordinate data and hence there was a need for new methodologies. 
Early works mainly utilised exploratory data analysis and visualisation techniques to blend the results from different studies \citep{Fox1998} and it was not until the early 2000's that the first methods for CBMA were proposed \citep{Fox1997,Turkeltaub2001,Nielsen2002,Wager2003}. 
Since then, many new methods and modifications appeared in the neuroimaging \citep[to name a few]{Laird2005b,Wager2007,Radua2009,Turkeltaub2012,Caspers2014} as well as the statistics \citep{Kang2011,Yue2012,Kang2014,Montagna2016} literature.

All of these methods share the same goal: to identify areas of the human brain that show consistent activation across studies. 
The different approaches broadly fall into two main categories: \textit{kernel-based} and \textit{model-based} methods. 
In what follows, we present the most widely used methods in both categories.
We start by setting the notation used throughout the manuscript.

A typical CBMA dataset consists of a list of foci from $I$ independent studies. 
Each study $i$, $i=1,\dots,I$ comes with a set of 3-dimensional coordinates $\mathbf{x}_{ik}\in\mathcal{B}$, where $\mathcal{B}\subset\mathbb{R}^3$ is the standard atlas space and $k$ indexes the multiple foci for a particular study. 
Table \ref{tab:reviewdata} is part of a real dataset from a meta-analysis of emotion studies that will be analysed for the purposes of this review. 
In this example, $\mathbf{x}_{52}$ would correspond to the second foci ($\left[-34,52,8\right]^{\mathrm{T}}$) in the fifth study (Baker 1997, happy). 
We denote $\mathbf{x}_i=\bigcup_{k}{\mathbf{x}_{ik}}$, the complete set of foci reported in study $i$. 
Note that some of the studies (e.g.\ Damasio 2000, fear and Damasio 2000, anger) are obtained from the same experiment; we treat these studies as independent following the standard conventions in the field. 
Finally, we will denote as $v=\left[v_x,v_y,v_z\right]\in\mathcal{B}$ the center location of a particular voxel in the brain atlas, $v=1,\dots,V$.

\begin{table}[h]\begin{center}
\caption{A subset of data from a meta-analysis study of emotions.}\label{tab:reviewdata}
\begin{tabular}{ccccccc}
\textbf{Author}&\textbf{Year}&\textbf{Emotion}&\textbf{X}&\textbf{Y}&\textbf{Z}&\textbf{Participants}\\\hline
Damasio &2000&fear&-10&-62&-17&23 \\
 &&&-1&-66&-1&23\\
 &&&34&3&32&23\\
Damasio &2000&anger&-2&-29&-12&23\\
Philips &2004&disgust&4&-20&15&8\\
 &&&7&-17&9&8\\
 &&&4&-63&26&8\\
Baker &1997&sad&36&20&-8&11\\
 &&&-44&32&-8&11\\
Baker&1997&happy&-26&28&0&11\\
&&&-34&52&8&11\\
Williams &2005&anger&7&31&28&13\\
 &&&7&28&-7&13\\
$\dots$&$\dots$&$\dots$&$\dots$&$\dots$&$\dots$&$\dots$\\ \hline
\end{tabular}
\end{center}\end{table}

\subsection{Kernel-based methods}
The most widely used kernel based methods are the {\em multilevel kernel density analysis} \citep[MKDA]{Wager2007}, the {\em activation likelihood estimation} \citep[ALE]{Eickhoff2012} and the {\em signed differential mapping} \citep[SDM]{Radua2012}. 
All these methods share the same rationale. 
Briefly, one starts by creating focus maps: that is, full brain images obtained through smoothing of reported activations with a spatial kernel. 
Obviously, there are as many focus maps as the total number of foci. 
Secondly, the focus maps corresponding to a particular study are combined to create the study-specific maps. 
These per-study images are subsequently combined into a single image that represents the evidence for consistent activation (clustering). 
Significance of these images is assessed with a Monte Carlo test under the null hypothesis of complete spatial randomness. 
We now discuss MKDA, ALE and SDM in detail.

\subsubsection{Multilevel kernel density analysis}
First introduced by \citet{Wager2003}, MKDA was modified to its current version by \citet{Wager2007}. 
To obtain the focus maps, $M_{ik}$, one places a sphere of unit intensity and radius $r$ centred at each focus:
\begin{equation}
M_{ik}(v) = \mathbf{1}_{\{ d(v,\mathbf{x}_{ik})\leq r \}}
\end{equation}
where $d(\cdot,\cdot)$ stands for the Euclidian distance. 
The study specific images, $M_i$, are then obtained by applying the maximum operator to the focus maps of the study. 
The procedure can be expressed by the following formula:
\begin{equation}\label{mkda1}
M_i(v)=
\begin{cases}1,&\exists k\hspace{0.1in} \mbox{s.t.} \hspace{0.1in} d(v,\mathbf{x}_{ik})\leq r\\ 0, &\mbox{otherwise}\\   \end{cases}.
\end{equation}
We call $M_i$ the {\em comparison indicator maps}. 
A value of 1 means that there is activation within distance $r$ of a given location. 
\citet{Wager2004} suggest giving $r$ a value of $10$ or $15$mm. 
The MKDA statistic image $m$ is given as a weighted combination of $M_i$:
\begin{equation}\label{mkda2}
m(v)=\frac{1}{\sum_{i}{w_i}}\sum_{i=1}^{I}{w_iM_i(v)}
\end{equation}
The weights are usually chosen to be proportional to the number of participants in each study thus allowing for studies with larger sample size to contribute more to the value of the statistic. 
If the weights are all set to 1 then $m(v)$ denotes the proportion of studies that reported activation within distance $r$ to $v$. 
Large values of $m\left(v\right)$ suggest systematic clustering of foci around its location. 

The distribution of the MKDA statistic does not have a closed form and thus Monte Carlo testing is used to assess significance. 
Multiple synthetic datasets are created by uniformly drawing peak locations from $\mathcal{B}$, keeping the original number of foci per study fixed. 
The $m$ statistic map is calculated for these datasets and the maximum value is saved at each replicate. 
This produces a sample of the maximal statistic under the null hypothesis of random foci allocation. The sample is then used to obtain FWE corrected $p$-values \citep{Kober2008} as suggested by \citep{Nichols2002}. 
Recently, \citet{Costafreda2009} derived a parametric significance test based on the properties of the spatial Poisson process. 
For applications of MKDA on real data see \citet{Etkin2007} and \citet{Kober2008}.

\subsubsection{Activation likelihood estimation}
The idea behind ALE is to model the probability that a voxel $v$ is the true location of a reported focus $\bx_{ik}$ as a discretised (over voxels) and truncated (outside the brain) Gaussian distribution centered around the reported location \citep{Turkeltaub2001}. 
For a certain study $i$, let $L_{ik}$ be the map based on a single focus $\mathbf{x}_{ik}$,
\begin{equation}
L_{ik}(v) = c\phi_3(v\mid \mathbf{x}_{ik},\sigma_i^2\mathbf{I}),
\end{equation}
where $\phi_3(\mathbf{x}; \boldsymbol\mu,\boldsymbol\Sigma)$ is the density of a three dimensional Gaussian distribution with mean $\boldsymbol\mu$ and covariance matrix $\boldsymbol\Sigma$ evaluated at $\mathbf{x}\in\mathbb{R}^3$, $\mathbf{I}$ is the identity matrix, and $c$ is a normalising constant ensuring that the sum of $\phi_3(\cdot)$ over voxels is equal to one. 
Thus, $L_{ik}(v)$ represents the probability that voxel $v$ is the true location of focus $\mathbf{x}_{ik}$. 
The Gaussian kernel used for ALE is analogous to the uniform kernel used for MKDA, but assigns higher values to voxels closer to the foci. 
To determine $\sigma_i$, \citet{Eickhoff2009} created a mapping between the number of participants in each study, $n_i$, and the standard deviation $\sigma_i$, the intuition being that larger sample sizes lead to improved spatial precisions. 
However, the mapping is based on an empirical study consisting of 21 subjects and may be unsuitable for experimental paradigms other than the one used by the authors. 

The next step for the ALE algorithm is to combine the focus maps $L_{ik}$ into a single study map $L_i$. Early versions of the algorithm \citep{Turkeltaub2001,Eickhoff2009} constructed study maps as the probability that at least one focus is located at a given voxel $v$\footnote{This is achieved by taking $L_i(v)=1-\prod_{k}{\left(1-L_{ik}(v)\right)}$}. 
The drawback of this approach is that it leads to ALE scores that are heavily influenced by studies reporting several foci per region. 
To address this issue, \citet{Turkeltaub2012} proposed taking the maximum over focus maps as in MKDA, that is,
\begin{equation}
L_i(v) = \max_k{L_{ik}(v)}.
\end{equation}
We call $L_i$ the \textit{modelled activation map}. 
$L_i(v)$ quantifies the probability that the focus which is closest to $v$ is truly located at $v$.

The ALE statistic $\ell$ is then computed as:
\begin{equation}\label{ale}
\ell(v)=1-\prod_{i=1}^{I}\left(1-L_{i}(v)\right).
\end{equation} 
Expression \ref{ale} was originally adopted by \citet{Turkeltaub2001} and  represents the probability that at least one of the closest activations is truly located in voxel $v$.

The Monte Carlo significance test of ALE is equivalent yet slightly different to the one of MKDA. 
In particular, \citet{Eickhoff2009} maintain that the spatial arrangement of activations (focus-to-focus distances) within a study must be preserved when generating synthetic studies, so that convergence across studies (rather than convergence over individual foci) is assessed. 
Hence for location $v$, multiple realisations of the ALE statistic $\ell^*(v)$ are created by sampling each activation map from a random location. 
The null ALE are then obtained as
\begin{equation}\label{alemc}
\ell^*(v) = 1-\prod_i(1-L_i(v^*)),
\end{equation}
where $v^*$ is drawn uniformly from all possible brain locations. 
By using \ref{alemc}, \citet{Eickhoff2012} showed that it is possible to enumerate exhaustively all the possible outcomes of $\ell$, directly obtaining the exact marginal null distribution of the ALE statistic without Monte Carlo. 
The null can be used to compute (uncorrected) $p$-values and from these FDR-corrected $p$-values \citep{Laird2005b}. 
FWE-corrected voxel-wise and cluster size, however, require the Monte Carlo simulation of null distribution of the maximum of $\ell$ \citep{Eickhoff2012}. 
\citet{Eickhoff2016} advocated FWE-corrected voxel-wise inference; they found, using simulated data, that this is less susceptible to ``spurious'' findings compared to voxel-wise FDR correction.

ALE has been used for several analyses including those in \citet{Delvecchio2012} and \citet{Konova2013}. 
ALE's popularity has been supported by the availability of software like \textit{GingerALE}, a broadly used, freely available implementation of the ALE algorithm. 
However, a recent contribution by \citet{Eickhoff2017} reported the presence of several errors in the ALE code. 
In particular, errors were traced in the voxel-wise FDR and cluster-wise FWE correction procedures for multiple testing, and have resulted in thresholds that were more liberal compared to those specified by users. 
For more details, see \citet{Eickhoff2017}. 
As of April 2017, we remark that these issues have been fixed.

\subsubsection{Signed differential mapping}
SDM \citep{Radua2009} is a relatively new method that borrows several characteristics from both MKDA and ALE. 
The novelty of the method lies in incorporating the $T$ statistic values (when available) from the original studies. 
To make this point clear, imagine that a study investigates brain activation caused by a given task; in some regions of the brain hyperactivation will be observed while in others there will be underactivation. 
In both cases, significant values of the $T$ statistic will be recorded; these values will be large and positive in the first case and large and negative values in the second. 
This case is particularly interesting when difference in activation between tasks is being investigated.

Assume that $T_{ik}$ is the reported $T$ value for focus $\mathbf{x}_{ik}$. SDM will generate focus maps $S_{ik}$ as:
\begin{equation}
S_{ik}(v) =\mathrm{sign}\left(T_{ik}\right) \exp{\left(-\frac{d(v,\mathbf{x}_{ik})^2}{2\sigma^2}\right)}.
\end{equation}
Similarly to ALE, SDM employs a Gaussian kernel, and the authors suggest using a standard deviation of approximately 25mm \citep{Radua2009}. 
The study maps are then:
\begin{equation}
S_i(v)=  \begin{cases}-1,&\sum_{k}{S_{ik}(v)}\leq -1\\\sum_{k}{S_{ik}(v)},&-1\leq \sum_{k}{S_{ik}(v)}\leq 1\\ 1,&\sum_{k}{S_{ik}(v)}\geq 1\\    \end{cases}.
\end{equation} 
That is, the study map is obtained as the sum of the corresponding focus maps, but is forced to lie within the interval $[-1,1]$ in the same way the MKDA study maps $M_i$ are given a maximum value of 1. 
Finally, the SDM statistic image, $s$, is calculated as the weighted mean of the study specific maps at each voxel:
\begin{equation}
s(v)=\frac{1}{\sum_i{w_i}}\sum_i{w_iS_i(v)}.
\end{equation}
Weights are once again proportional to number of participants in the study. 
Since the method averages both positive and negative findings, voxels that show contradicting results will not appear as significant. 
Inference is based on the same Monte Carlo scheme of MKDA, and thresholding is done either by setting a highly conservative rejection point ($p<0.001$) or controlling the FDR \citep{Radua2009}. 

In a newer version of the algorithm, the authors use the $T_{ik}$ values to reconstruct the original $T$ statistic images. 
That way, it is possible to incorporate both CBMA and IBMA data in the same analysis. 
For more details, see \citet{Radua2012}. 
The last contribution made on SDM lies in the use of anisotropic kernels in the analysis \citep{Radua2014}. 
Anisotropy can be easily incorporated in MKDA and ALE but its superiority to the current practice of using isotropic kernels is only based on empirical findings, thus it should be further investigated. 
Published work utilising SDM for the analyses includes \citet{Richlan2011} and \citet{FusarPoli2012}.

\subsection{Model-based Methods}\label{sec:modelbased}
Recently there has been growing interest in the development of model-based methodologies to address some of the limitations of kernel-based methods. 
These methods use ideas from spatial statistics to develop stochastic models for the analysis of foci. 
Unfortunately the literature on model based methods is still very limited thus our review will be almost exhaustive. 
In particular, we will outline the {\em Bayesian hierarchical cluster process} model of \citet[BHICP]{Kang2011}, the {\em spatial binary regression} model of \citet{Yue2012}, the {\em hierarchical Poisson/Gamma random field} model of \citet{Kang2014}, and the {\em spatial Bayesian latent factor regression} model of \citet{Montagna2016}.  
In all cases analyses are performed under the Bayesian paradigm and thus inferences are based on posterior distributions for each model's parameters.  

Some of the methods reviewed here are build upon spatial point processes theory. Spatial point processes are random sets of points in the $d$-dimensional Euclidian space. 
A detailed description of the theory behind point processes is beyond the scope of this review, so we refer the reader to \citet{Moller2004} and \citet{Illian2008} for details and applications. We now proceed to describe the details of model-based methods.

\subsubsection{A Bayesian hierarchical independent cluster process model (BHICP)}\noindent 
\citet{Kang2011} proposed a hierarchical model based on an independent cluster process to describe the mechanism generating the foci. 
The model is structured into 3 levels, of which the lowest level, level $1$, contains the observations (foci), while higher levels describe the study and population structure respectively. 
The distinction between singletons and multiple foci is incorporated into the model. 
In the outline of the model below, we occasionally suppress the $k$ index so that $\textbf{x}_i$ is the set of foci reported in study $i$, i.e.\ $\textbf{x}_i=\bigcup_k{\mathbf{x}_{ik}}$.

\begin{figure}[htp]
\centerline{
\includegraphics[scale=0.85]{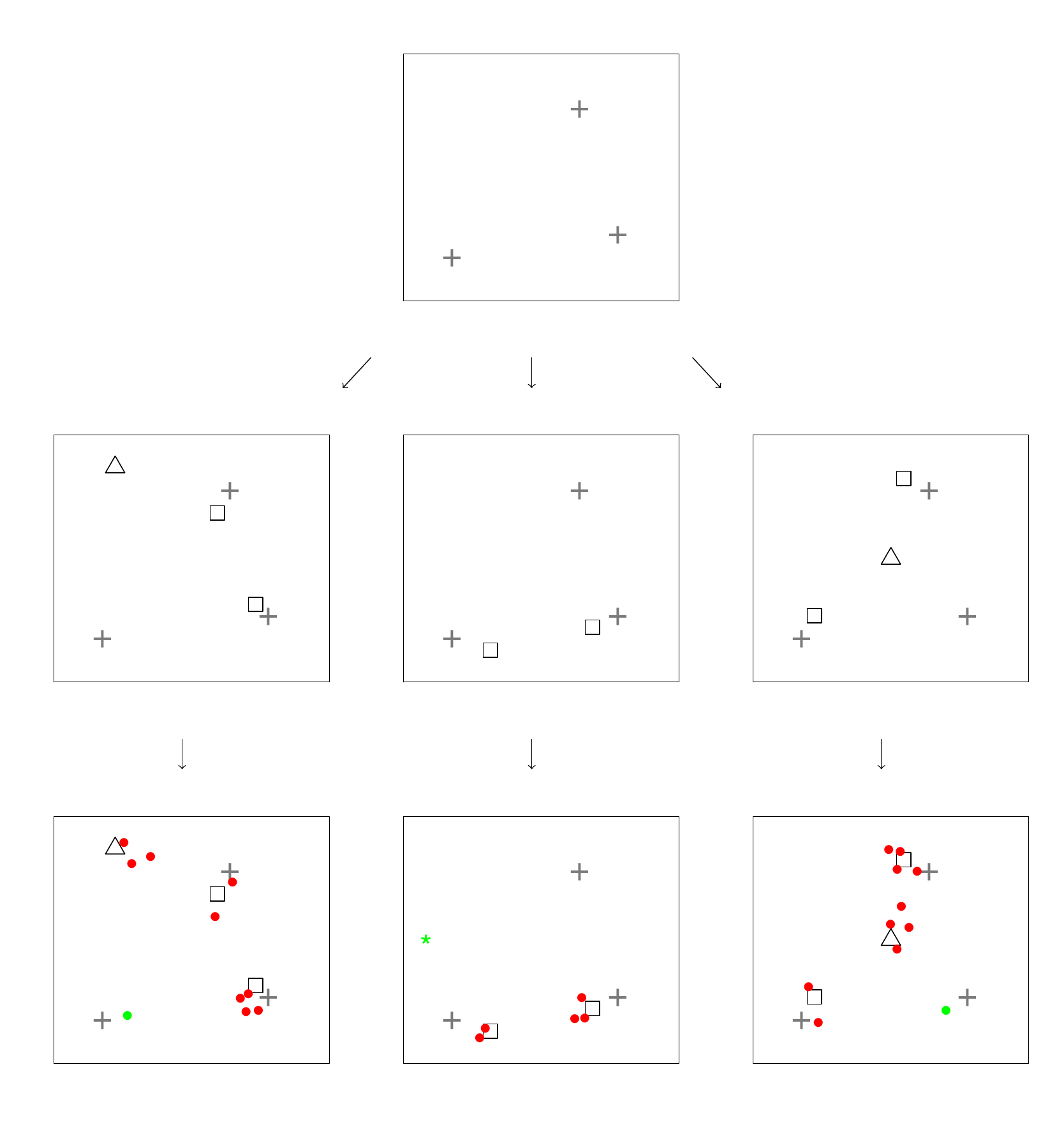}}
\caption{Realisation of the BHICP model for 3 studies. At level 3 (top) latent population centres (grey, $\mathbf{z}$) lie. At level 2 (middle) we have centres of multiple foci (black). These come either directly from population centres (squares, $\mathbf{y}_i$) or from background noise (triangles, $\mathbf{y}_{i\emptyset}$). Level 1 (bottom) contains the data ($\mathbf{x}_i$). These are multiple (red, $\mathbf{x}_i^1$) or singleton (green, $\mathbf{x}_i^{0}$) foci. Singletons come either directly from population centres (dots, $\mathbf{x}_{i\xi}^{0}$) or from a background Poisson process (asterisks, $\mathbf{x}_{i\emptyset}^{0}$).}
\label{b}
\end{figure}

Figure \ref{b} provides a graphical representation of the model. 
At level 1, we have the foci (Fig. \ref{b} bottom, coloured circles). 
We denote with $\textbf{X}_i$ the underlying process generating the observations $\textbf{x}_i$ in each study. 
As discussed in Section \ref{sec:neuroback}, we can have both singleton and multiple foci. 
Thus, process $\textbf{X}_i$ consists of two mechanisms, one generating the multiple foci (Fig.\ \ref{b}, red circles) and one that is giving the singleton foci (Fig.\ \ref{b} bottom, green circles): $\textbf{X}_i=\textbf{X}_i^1\cup\textbf{X}_i^0$. 

Multiple foci $\textbf{X}_i^1$ can be viewed as an independent cluster process of points centered around study activation centers $\mathbf{y}_i$. 
In particular, for every study center $\psi\in\mathbf{y}_i$ we have a process $\mathbf{X}_{i\psi}^1$ of multiple foci, where total number of offsprings has a $\mathrm{Pois}(\eta_i)$ distribution. 
Conditional on the total number of offsprings, all the points $\chi\in\mathbf{X}_{i\psi}^1$ are normally distributed around $\psi$ with covariance $\mathbf{T}_\psi$, that is $\chi\sim\mathcal{N}\left(\psi,\mathbf{T}_{\psi}\right)$, for all $\chi\in\mathbf{X}_{i\psi}^{1}$.
In other words, each $\mathbf{X}_{i\psi}^1$ is a Poisson process defined on the brain with intensity given by:
\begin{equation}
\lambda_{i\psi}^{1}(\xi)=\eta_i\phi_3(\xi\mid \psi,\mathbf{T}_\psi), \quad \xi\in\mathcal{B}. 
\end{equation}
The overall observed pattern of multiple foci is then given as the union of all offsprings, i.e.\ $\textbf{X}_i^1=\bigcup_{\psi\in\textbf{y}_i}{\textbf{X}^1_{i\psi}}$, and therefore is also a Poisson process with intensity:
\begin{equation}
 \lambda_{i}^{1}(\xi)=\sum_{\psi}{\lambda^{1}_{i\psi}(\xi)}.
 \end{equation}

Singleton foci $\textbf{X}_i^0$ arise directly from the population center process $\mathbf{z}$. 
As with multiple foci, $\textbf{X}_i^0$ has a total of $\mathrm{Pois}(\theta_i)$ points which are normally distributed around the population center locations $\zeta\in\mathbf{z}$, i.e.\ $\chi\sim\mathcal{N}\left(\zeta,\boldsymbol\Sigma_{\zeta}\right) $ for all $\chi\in\mathbf{X}_{i\zeta}^{0}$ and covariance matrix $\boldsymbol\Sigma_\zeta$.
To add more flexibility, the model allows for some singletons to not cluster around any population center, say $\textbf{x}_{i\emptyset}$. 
These foci are assumed to arise from a Poisson process $\textbf{X}_{i\emptyset}$ of constant intensity $\epsilon_{1i}$:
\begin{equation}
\textbf{X}_{i\emptyset}\mid \epsilon_{1i}\sim\mathcal{PP}\left(\mathcal{B},\epsilon_{1i}\right).
\end{equation}
Overall, $\textbf{X}_i^0=\left(\bigcup_{\zeta\in\mathbf{z}}{\textbf{X}_{i\zeta}^0}\right)\cup\textbf{X}_{i\emptyset}$ are the singleton foci of a study. 
Hence, the intensity of $\mathbf{X}_i^0$ is:
\begin{equation}
\lambda_{i}^0(\xi)=\epsilon_{1i}+\theta_i\sum_{\zeta\in\mathbf{z}}\phi_3\left(\xi\mid\zeta,\boldsymbol\Sigma_\zeta\right),\quad \xi\in\mathcal{B}.	
\end{equation}
While kernel-based methods condition on the total number of activations, in the BHICP model the total number of foci is a Poisson random variable with mean $\int_\mathcal{B}\lambda_i(\xi)d\xi$, where $\lambda_i=\lambda_i^0+\lambda_i^1$. 

At level 2, we have the unobserved study activation centers $\textbf{y}_i$, which are the locations around which the multiple foci of a study cluster. 
The $\mathbf{y}_i$ are realisations of a point process $\textbf{Y}_i$, and may either cluster around the population centres $\textbf{z}$ (Fig.\ \ref{b} middle, squares) or appear in random locations across the brain (Fig. \ref{b} middle, triangles). 
To account for the former, clustered study centers $\textbf{Y}_{i\zeta}$ are introduced as sets of points normally distributed around population centers $\zeta \in \textbf{z}$, and with covariance matrix $\mathbf{\Sigma_\zeta}$. 
The total number of points in each $\textbf{Y}_{i\zeta}$, that is, the total number of clustered study centers for study $i$ is a Poisson random variable with mean $\kappa_i$.
As for the latter, noise study centers are modelled as a homogenous Poisson process $\textbf{Y}_{i\emptyset}$ with intensity $\epsilon_{2i}$. 
Overall, $\textbf{Y}_i=\left(\bigcup_{\zeta\in\textbf{z}}{\textbf{Y}_{i\zeta}}\right)\cup\textbf{Y}_{i\emptyset}$ is Poisson point process with intensity given by:
\begin{equation}
	\rho_i(\xi)=\epsilon_{2i}+\kappa_i\sum_{\zeta\in\mathbf{z}}{\phi_3\left(\xi\mid\zeta,\boldsymbol\Sigma_\zeta\right)}, \quad\xi\in\mathcal{B}.
\end{equation}

At the highest level (level 3), we have the population activation centres (Fig.\ \ref{b} top, gray crosses). 
These are unobserved realisations $\textbf{z}$ of an \textit{a priori} homogenous Poisson process $\textbf{Z}$ of intensity $\epsilon_3$:
\begin{equation}
\textbf{Z}\mid \epsilon_{3}\sim\mathcal{PP}\left(\mathcal{B},\epsilon_{3}\right)
\end{equation}
Population centres are the locations around which study activation centers and singleton foci scatter. 
As such, they can be viewed as locations in the brain where an overall population effect exists. 

The BHICP can be viewed as a random effects model as it allows for both within-study and between-study variability. 
Samples from the posterior distributions are obtained via MCMC. 
Several interesting quantities can be inferred upon such as regions of consistent activations (through the posterior distribution of populations centers), the uncertainty in the location of study centers around the population centers (through $\mathbf{\Sigma}_\zeta$) and the variability of the foci within studies (through $\mathbf{\Psi}_\xi$).

\subsubsection{A Bayesian spatially adaptive binary regression model}
\citet{Yue2012} use spatial logistic regression for a meta-analysis of emotion studies. 
For study $i$ and voxel $v$, let $y_i\left(v\right)$ be the binary outcome defined as:
\begin{equation}
y_i(v)=\begin{cases}
1&\mbox{at least one focus at voxel $v$}\\
0&\mbox{no foci at voxel $v$}\\
\end{cases}.
\end{equation}
Note that the binary study images $\left\{y_i\left(v\right)\right\}_{v=1}^V$ are identical to the MKDA study maps $M_i\left(v\right)$.
Logistic regression can be used to model the probability that a voxel is reported as a focus, $p_i(v)=\mathbb{P}\left(y_i\left(v\right)=1\right)$. 
It is assumed that:
\begin{equation}
p_i(v)=H\left(z(v)\right),
\end{equation} 
where $H\left(\cdot\right)$ is the link function. 
The authors use the standard probit and logit link functions. 

Spatial correlation is induced through the prior on $\left\{z\left(v\right)\right\}_{v=1}^{V}$. 
In particular, we assume that the process $z\left(v\right)$ is an adaptive Gaussian Markov random field \citep[aGMRF]{Yue2010}. 
The aGMRF model defines the conditional distribution of $z\left(v\right)$ through a specific dependence with neighbouring voxels.
A significant merit of the method is the inclusion of a local smoothness parameter $\gamma\left(v\right)$ for the aGMRF.
This allows the method to automatically choose the amount of smoothing required depending on the amount of information available. 

Authors further introduce a process $\psi\left(v\right)$, an indicator of whether the outcome variable $y_i\left(v\right)$ is miscoded; the case $\psi\left(v\right)=1$ can either refer to both false positives, voxels that were falsely found as activated, and false negatives, voxels that were not reported as foci even though they were activated.
The process $\psi\left(v\right)$ is not observed and hence is estimated along with the remaining model parameters. 

Posterior probabilities of activation at each voxel are obtained through an auxiliary variable MCMC algorithm. 
Voxels with high posterior probabilities of being reported as foci are more likely to show an effect. 
A potential drawback of the method is that it can be currently applied only in two dimensions. 
In three dimensions, the value of one of the axes is held fixed, for example $z=c$, while the model is fitted for all available observations of the form $\mathbf{x}_{ik}=\left[ x1_{ik},x2_{ik},c\right]$. 
Authors however, maintain that extending the model to three dimensions is possible.

\subsubsection{A hierarchical Poisson/Gamma random field model (HPGRF)}
A neuroimaging meta-analysis will typically consider several subtypes of tasks. 
For example, a meta-analysis of emotion may classify the studies according to experiments on ``happiness", ``sadness", ``pain", etc. 
Yet, the methods described previously are for a single homogeneous group of studies. \citet{Kang2014} propose a model that models each type of foci separately, allowing simultaneously for dependence between the $J$ different types.

Let $\mathbf{x}_{ij}$ be the set of foci reported by study $i$ for task type $j$. 
Suppose that $\mathbf{x}_{ij}$ are realisations of a Cox Process $\mathbf{X}_j$ driven by a random intensity measure $\mathbf{\Lambda}_j(d\xi)$. 
Conditional on $\mathbf{\Lambda}_j(d\xi)$, $\mathbf{X}_j$ are Poisson processes on the brain $\cal B$:
\begin{equation}\label{pgrf1}
\mathbf{X}_{j}\mid\mathbf{\Lambda}_j(d\xi)\sim \mathcal{PP}(\mathcal{B},\mathbf{\Lambda}_j(d\xi)).
\end{equation}

In other words, we have $J$ underlying Cox processes, each one contributing a specific type of foci in some/all of the studies. 
The intensity measures $\mathbf{\Lambda}_j(d\xi)$ arise from a convolution of a finite kernel measure $\mathbf{K}_j(d\xi,\zeta)$ and a Gamma Random Field $\mathbf{G}_j(d\zeta)$:
\begin{equation}\label{pgrf2}
\mathbf{\Lambda}_j(d\xi)=\int_{\mathcal{B}}{\mathbf{K}_j(d\xi,\zeta)\mathbf{G}_j(d\zeta)}.
\end{equation}

The model arising from (\ref{pgrf1})-(\ref{pgrf2}) is similar to the Poisson/Gamma random field model of \citet{Wolpert1998}, who first introduced the idea of convolving a Gamma random field with a Poisson process. 
To introduce dependence between the different tasks, it is assumed that $\mathbf{G}_j(d\zeta)$ are independent realisations of a Gamma random field with common shape measure $\mathbf{G}_0(d\zeta)$ and inverse scale parameter $\beta$:
\begin{equation}
\mathbf{G}_j(d\zeta)\sim \mathcal{GRF}(\mathbf{G}_0(d\zeta),\beta).
\end{equation}

Again, $\mathbf{G}_0(d\zeta)$ is a Gamma random field:
\begin{equation}
\mathbf{G}_0(d\zeta)\sim \mathcal{GRF}(\mathbf\alpha(d\zeta),\beta_0).
\end{equation}

An MCMC scheme is used for posterior computation. The HPGRF model allows for the detection of overall effects based on the posterior intensity $\mathbf{G}_0(d\zeta)$ or task-specific effects based on $\mathbf{\Lambda}_j(d\xi)$. 
Inference on types with fewer observations can be done by borrowing information from the remaining types through correlation under the common base intensity $\mathbf{G}_0(d\zeta)$. 
A significant benefit of the model is that it requires the specification of very few hyperparameters.

\subsubsection{A spatial Bayesian latent factor regression model}
In a recent contribution, \citet{Montagna2016} generalise the model proposed in \citet{Montagna2012} to the case where observations are spatial point patterns reported from different neuroimaging studies.
The authors regard the foci from each study $\bx_i$ as a Cox process $\bX_i$ driven by a non-negative random intensity function $\mu_i$:
	\begin{equation}
	\bX_i \mid \mu_i \sim \mathcal{PP}(\mathcal{B}, \mu_i).
	\end{equation}

\citet{Montagna2016} consider a functional representation for the (log) intensity function and write $\log \mu_i$ in terms of a collection of basis functions:  
	\begin{equation}\label{logintensity}
	\log \mu_i(\bnu) = \sum_{m = 1}^p \theta_{im} b_m(\bnu) = \bbb(\bnu)^{\top}\btheta_i.
	\end{equation}

This specification implies that $\log \mu_i$ belongs to the span of a (fixed) set of basis functions, $\{b_m(\cdot)\}_{m = 1}^p$, with $\btheta_i$ denoting a vector of study-specific coefficients. 
For a discussion on the choice of the bases (e.g., B-splines, Gaussian kernels, etc.) and their number $p$ we refer to \citet{Montagna2016}. 
A low dimensional representation of $\log \mu_i$ is achieved by placing a sparse latent factor model \citep{Arminger1998} on the basis coefficients:
		\begin{equation}
		\btheta_i = \bLambda \bbeta_i + \bzeta_i, \quad \text{with} \quad \bzeta_i \sim N_p(0, \bSigma)
		\end{equation}
where $\btheta_i = [\theta_{i1}, \dots, \theta_{ip}]^\top$, $\bLambda$ is a $p \times k$ factor loading matrix with $k\ll p$, $\bbeta_i = (\eta_{i1}, \dots, \eta_{1k})^\top$ is a vector of latent factors for study $i$, and $\bzeta_i = (\zeta_{i1}, \dots, \zeta_{ip})^\top$ is a residual vector that is independent with the other variables in the model and is normally distributed with mean zero and diagonal covariance matrix $\bSigma = \text{diag}(\sigma_1^2, \dots, \sigma_p^2)$. \\
\indent Two attractive features of this approach are the ability to accommodate covariate information ({\it meta-regression}) and perform {\it reverse inference}. Both goals are achieved by putting the low dimensional vectors of latent factors $\bbeta_1, \dots, \bbeta_n$ in any flexible joint model with other variables of interest. For example, information from covariates $\bZ_i$ can be incorporated through a simple linear model: 
	\begin{equation}
	\bbeta_i = \boldsymbol{\beta}^\top \bZ_i + \bDelta_i, \quad \text{with} \quad \bDelta_i \sim N_k(0, \bI),
	\end{equation}
where $\boldsymbol{\beta}$ is a $r \times k$ matrix of unknown coefficients, and $r$ denotes the dimension of $\bZ_i$. 

Finally, reverse inference refers to inferring which cognitive process or task generated an observed activation in a certain brain region. In mathematical terms, it corresponds to estimating Pr[Task $\vert$ Activation]. Suppose for simplicity that the meta-analysis dataset consists of studies that can be categorized as either type A (e.g., happy) or type B (e.g., sad). Let $y_i$ denote the study type, with 
 \[ y_i = \left\{ 
  \begin{array}{l l}
    1 & \quad \text{if study $i$ is type A}\\
    0 & \quad \text{if study $i$ is type B}.
  \end{array} \right.\]
The interest is in estimating the probability that newly observed point pattern data arose from either a type A or B experiment. Because the study type can be represented as a binary response, the authors build a probit model for study type and predict the posterior probability that a new point pattern data arose from either type. Specifically, they model $p_{y_i} = $ Pr$(y_i = 1 \vert \alpha, \bgamma, \bbeta_i) = \Phi(\alpha + \bgamma^\top \bbeta_i)$, where $\Phi(\cdot)$ denotes the standard normal distribution function. Parameter $\alpha$ can be interpreted as the baseline probability that study $i$ is of type A, and $\bgamma^{\top}\bbeta_i$ accounts for study-specific random deviations. Notice that the latent factors $\bbeta_i$ are used as a vehicle to link the random intensities (thus, the foci) to the study-type. We remark that the probit model can be easily replaced by an appropriate predictive model for categorical, nominal, or continuous study features.

\section{Evaluation of existing methods}\label{sec:evaluations}
One of the aims of this paper is to evaluate CBMA methods. 
A head-to-head comparison of existing methodologies is unfeasible, because the statistics described earlier have very different interpretations. 
Instead, we examine some characteristics of CBMA methods that show the drawbacks and merits of each. 
In what follows, we focus on the comparison between kernel-based and model-based methods. 
In Section \ref{sec:reviewale} we conduct a series of simulations to study the sensitivity properties of the ALE algorithm that we think characterise other kernel-based methods as well. 
In Section \ref{sec:reviewreal}, we apply the methods for which available software exist on a real dataset and compare the outputs. 
Finally, in Section \ref{sec:reviewdiscussion} we proceed to a discussion.

\subsection{ALE simulation study}\label{sec:reviewale}
Even though kernel-based methods have been extensively used for the analysis of neuroimaging data, their power properties have not been thoroughly investigated on synthetic datasets. 
We perform a simulation study to assess the power properties of the ALE method. 
In particular, we want to assess how the power of the algorithm evolves with respect to the number of studies in the meta-analysis and whether  the method is robust to the inclusion of low quality studies. 
We choose ALE for three main reasons. 
Firstly, ALE is currently the most broadly used method for CBMA (based on a PubMed search for ALE, MKDA and SDM). 
Secondly, a recent review of kernel-based methods \citep{Radua2012b} reported that the three kernel-based methods provide qualitatively similar results, thus we expect that our findings are indicative of MKDA and SDM methods as well. 
Finally, we strongly believe that the current version of ALE \citep{Eickhoff2012} provides the best approximation to the Monte Carlo test null distribution upon which inference is based.  

We create meta-analytic datasets based on the following setup. 
Each simulated dataset consists of $I$ studies; of these,  $Ip$ are valid while the rest $I(1-p)$ are noise, $0\leq p\leq1$. 
For the valid studies, we assume there exist 8 population centers around which foci cluster. 
For each center, a valid study reports no foci with probability 0.35 (that is, each center is detected with probability 0.65), a singleton focus with probability 0.5, two multiple foci with probability 0.1 and three multiple foci with probability 0.05.
Hence, the expected number of foci per valid study is 6.8, similar to the average number of foci in the application of Section \ref{sec:reviewreal}. 
The foci are drawn from a three dimensional Gaussian distribution centered at the corresponding population center. 
As for the noise studies, we simply sample foci uniformly from the brain mask. 
The expected number of foci for the noise studies is the same as for valid studies. 
Our setup for a power analysis of the ALE algorithm is similar to the one in \citet{Eickhoff2016}. Nevertheless, their study involves a single activated region, and therefore fewer measures of power are considered. Further, their data generating mechanism is different to the one employed in this work.


As sample size, we consider $I$ equal to $20,40,60,80,100$ and $120$. 
For a given $I$, we successively set $p=0,0.05,0.10,0.15,...,0.95,1$. For each distinct combination of $I$ and $p$, we create $B=1,000$ datasets as described above, and apply the ALE algorithm \citep{Eickhoff2012} to each dataset. 
The normal kernel standard deviation is set to $\sigma=4$mm, constant across studies; the value is chosen according to the sample size-standard deviation mapping used by \textit{GingerALE}, and is used for studies with $\approx 12$ participants. 
We use an $\alpha=0.05$ FDR-corrected threshold to assess significance of the ALE statistic images. 
The following power-related quantities are recorded: 1) the probability that at least one of the 8 population centers is detected; 2) the probability that all 8 centers are detected; 3) the mean number of centers detected in 1,000 runs; 4) the mean voxel-wise true positive rate, where ``truly'' active voxels are defined by the 95\% probability spheres around the population centers.

\begin{figure}[h]
\centering
\includegraphics[scale=1.0]{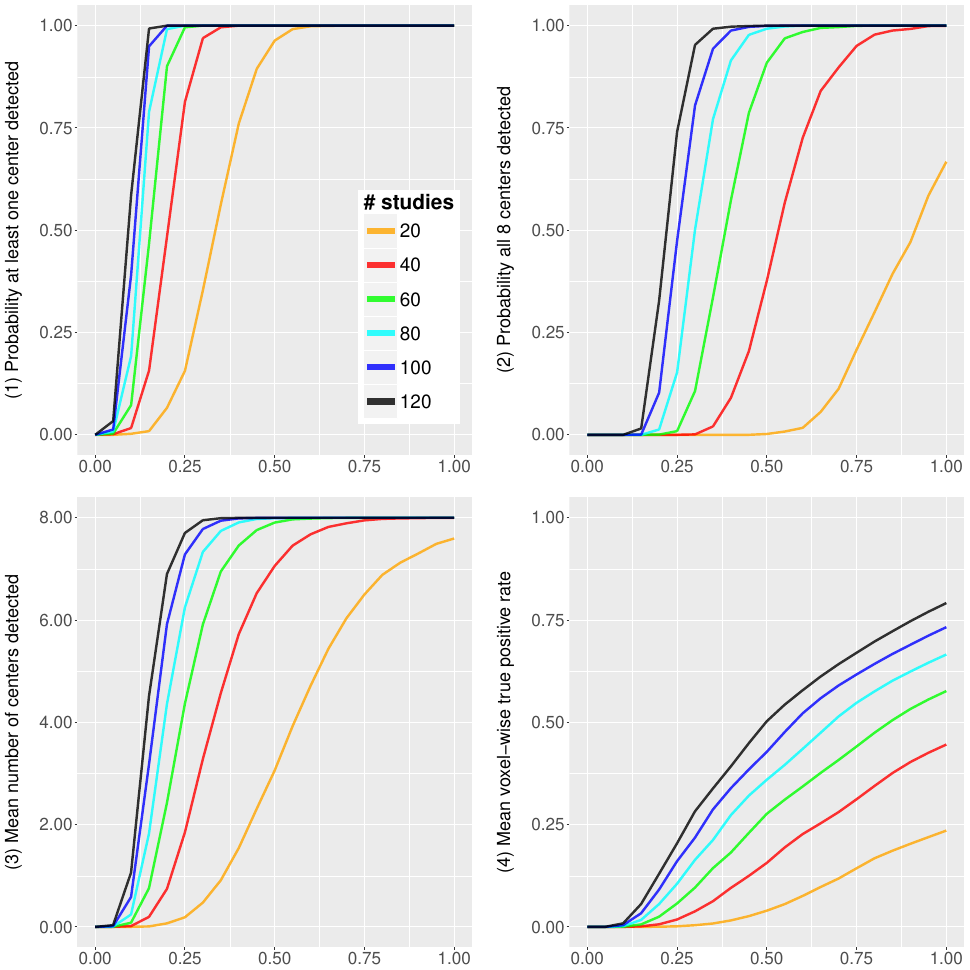}
\caption{Results of the simulation study. Power properties of the ALE algorithm are plotted against the proportion of valid studies $p$. Top left: probability that at least one center is detected. Top right: probability that all 8 centers are detected. Bottom left: mean number of centers detected. Bottom right: mean voxel-wise true positive rate.}
\label{fig:alesim1}
\end{figure}

\begin{figure}[h]
\centering
\includegraphics[scale=1.0]{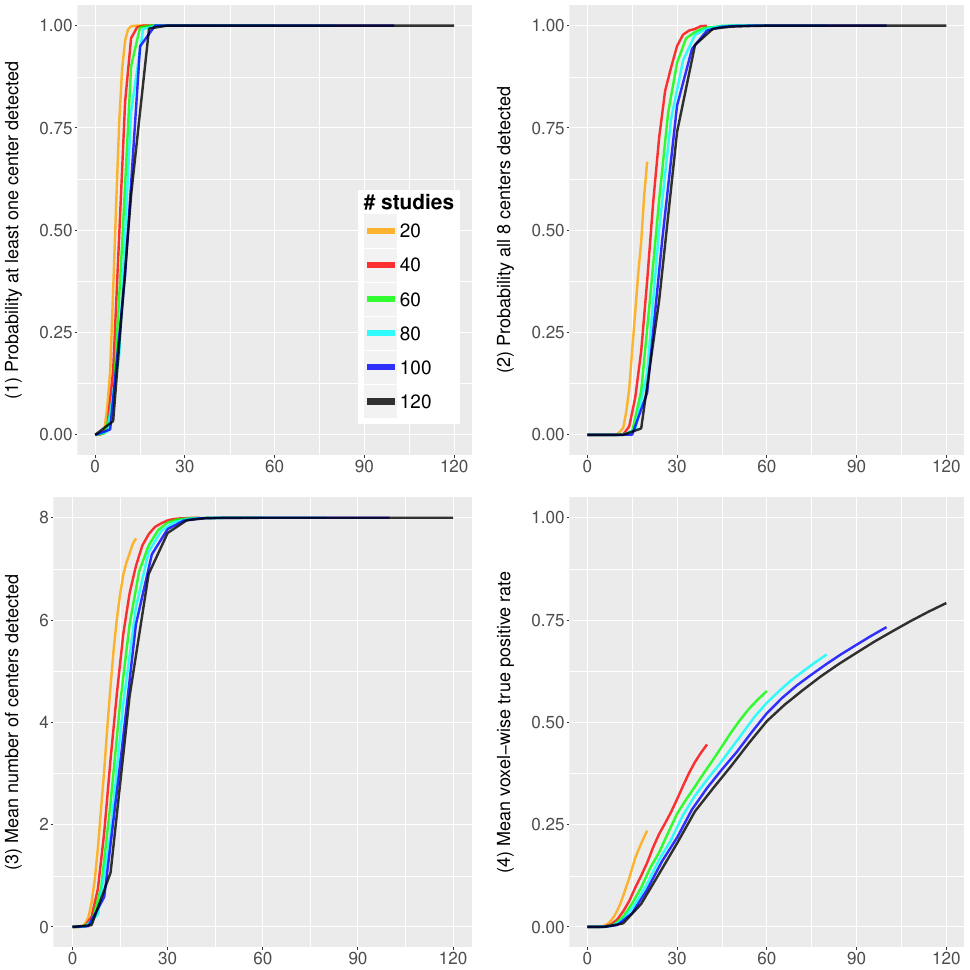}
\caption{Results of the simulation study. Power properties of the ALE algorithm are plotted against the total number of valid studies $Ip$. Top left: probability that at least one center is detected. Top right: probability that all 8 centers are detected. Bottom left: mean number of centers detected. Bottom right: mean voxel-wise true positive rate.}
\label{fig:alesim2}
\end{figure}

Our findings are summarised in Figure \ref{fig:alesim1} where quantities $1-4$ are plotted against the proportion of valid studies. 
One can observe that all 4 power measures increase monotonically to their maximal values of $1,1,8$ and $1$, respectively, as the number of studies grows. 
For a given proportion of valid studies, we observe that the bigger the total sample size is, the higher the power. 
For a fixed sample size, the power increases with the proportion of valid studies. 
Therefore, ALE is a consistent test. 
In Figure \ref{fig:alesim2} we plot quantities $1-4$ versus the total number of valid studies, that is, $Ip$ instead of $p$. 
We see that the curves for different $I$ tend to coincide. 
This is a key robustness property of the ALE algorithm, that is, adding pure-noise studies does not degrade power detection. 
Our results are consistent with the findings of \citet{Eickhoff2016}, who observe a similar behaviour of the ALE algorithm in their setup.

In Appendix \ref{sec:app2}, we repeat the simulation study above with both MKDA and SDM kernels, and using ALE's exact enumeration procedure to threshold the statistic images (5\%, FDR-corrected). 
This extension aims at investigating the effect of choosing three different kernels under a common inferential procedure.
We find that both MKDA and SDM kernels exhibit similar behaviour to ALE. 
For more details, see Appendix \ref{sec:app2}.  

\subsection{Analysis of a real dataset}\label{sec:reviewreal}
In this section, we perform a meta-analysis of emotion studies. 
The dataset we consider here was previously analysed by \citet{Kober2008}, \citet{Kang2011}, \citet{Yue2012} and \citet{Kang2014}, and is now publicly available\footnote{\url{https://osf.io/3vn9c/}}. Our goal here is not to extend the analyses in these papers, but rather to facilitate a comparison among the methods described in Section~\ref{sec:cbma} and, more generally, show how CBMA can be used to identify areas of consistent brain activation. Due to lack of software availability, however, we only apply MKDA\footnote{\url{https://github.com/canlab/Canlab_MKDA_MetaAnalysis}}, ALE\footnote{\url{http://www.brainmap.org/ale/} (RRID:SCR\_014921)}, SDM\footnote{\url{http://www.sdmproject.com/software/} (RRID:SCR\_002554)}, and the BHICP\footnote{\url{http://www-personal.umich.edu/~jiankang/software.html}} to the data.

The dataset consists of 164 experiments conducted between 1993 and 2005. 
Most of these experiments investigated multiple contrasts (2.7 contrasts per study on average), and each contrast involved a different emotion. 
It is also not unlikely that a given experiment used the same group of subjects to investigate multiple emotions, usually in the same scanning session. 
It is important to note that the CBMA methodologies reviewed in Section~\ref{sec:cbma} do not account for within-study correlation, thus effectively treat contrasts within each study (and across studies) as independent. 
Accommodating for within-study correlation may be infeasible due to the sparsity of the point patterns, as well as for computational considerations. 
While we recognise this may be a limitation, extending CBMA methods to account for within-study correlation is beyond the scope of this paper. 
Thus, we following standard neuroimaging convention and treat contrasts as independent hereafter.
Hence, our sample consists of $I=437$ contrasts and a total of 2478 foci, with an average of roughly 6 foci per contrast. 
Eight emotion types appear in the dataset: affective, anger, disgust, fear, happy, mixed, sad and surprise. 
In general, the goals of a meta-analysis are twofolds: 1) Identify consistent activation regions (aggregation of foci) across studies of the same emotion; and 2) Identify consistent activation regions across all emotion types. 
Since methods for which software is available can not accommodate a multi-type meta-analysis, we focus on the second objective here. 
This in turn enables us to pool all 437 contrasts together and work with a bigger sample size.

Note that 58 experiments used \textit{positron emission tomography} (PET) imaging instead of fMRI. 
Both PET and fMRI use blood flow to detect activation signals, but differ fundamentally in the way they exploit changes in blood flow to detect activations (for details on PET acquisition see e.g.\ \citep{Bailey2006}). Consequently, results may differ depending on which acquisition method was used. 
For example, it is well known that fMRI scanners provide better spatial resolution compared to PET scanners, but are also more sensitive to motion artefacts.   
In the emotions literature, there are both studies suggesting that results obtained from the two different modalities are similar \citep[for example]{Wager2008}, as well as studies revealing differences between the two (see, e.g., \citet{Costafreda2008}). 
While we chose to include both PET and fMRI studies in our analysis hereafter, we highlight the need for extending existing CBMA methodologies to account for nuisance effects such as different imaging modalities.

The simulation parameters are set as following. 
For MKDA, we use a kernel size of $r=10$mm, which is also the software default. 
A total of 10,000 Monte Carlo datasets are generated under the null hypothesis and used to the threshold the MKDA statistic image $m\left(v\right)$ at $\alpha=0.05$, FWE corrected.
ALE automatically assigns a kernel size for each study based on the total number of participants and uses the method of \citet{Eickhoff2012} to calculate the distribution of the statistic under the null hypothesis. 
The significance of the statistic image $\ell\left(v\right)$ is accessed with an FDR corrected $\alpha=0.05$ threshold.
For SDM we use an isotropic kernel of $20\mathrm{mm}$ since it is the software default and do 500 Monte Carlo randomisations.
For the BHICP, we use the same hyperparameter values as in \citet{Kang2011}, and run the MCMC for 120,000 iterations saving once every 100 iterations.
This results in a total sample size of 1200 posterior draws, of which we discard the first 200 as a burnin. 
The run length and burn-in are chosen following \citet{Kang2011}. 
We assess convergence visually by inspecting the traceplots of some of the model parameters. 
Further, we conduct a second run of the model initialised at different starting values. 
A comparison of the results from the two runs reveals no significant differences. 
For more details, see Appendix \ref{sec:app1}.
We now summarise the results. 

\begin{sidewaysfigure}[htp]
        \centering
        \includegraphics[scale=1.0]{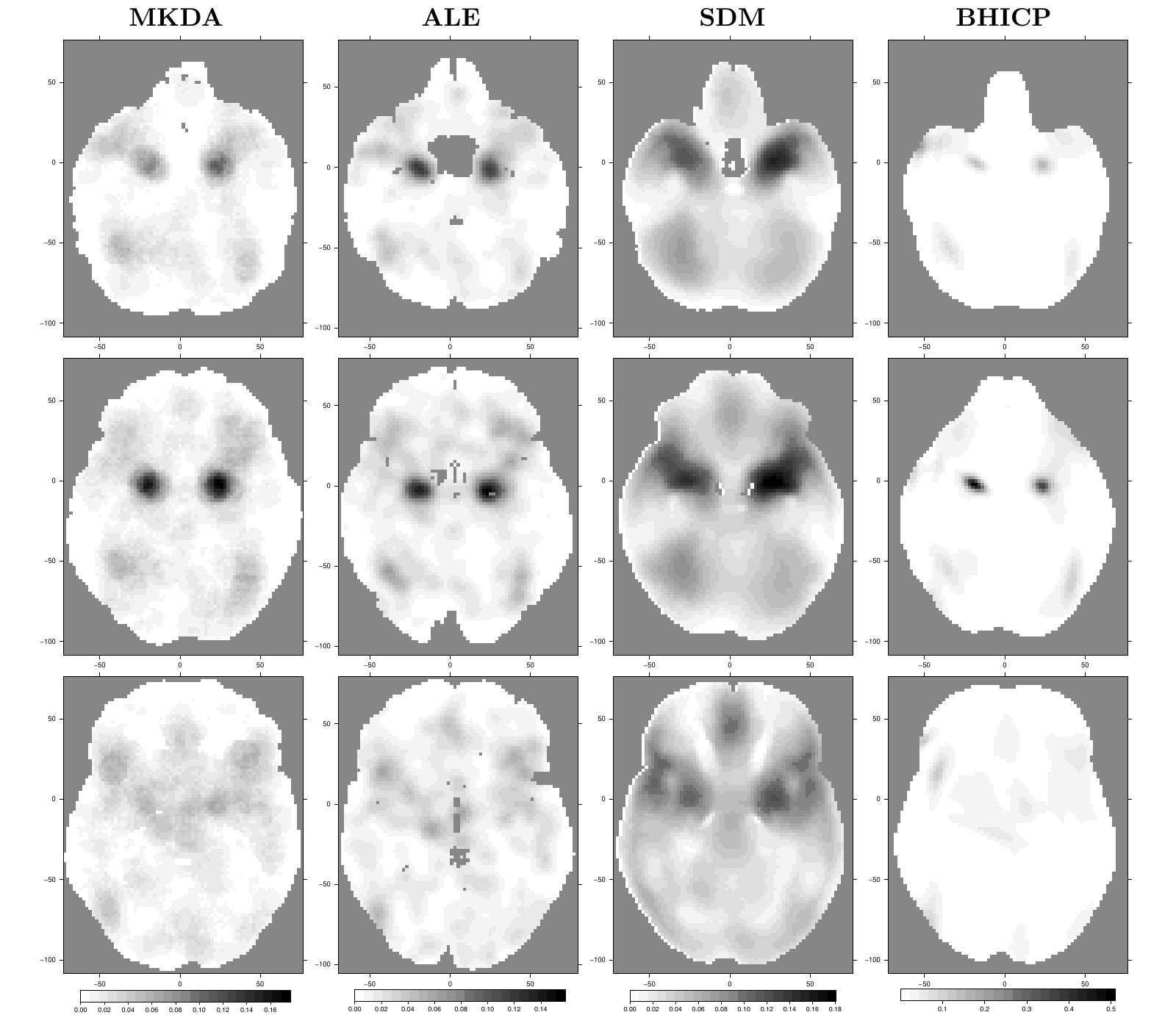}\\
        \vspace{-0.0in}
\caption{Qualitative comparison of the CBMA methods in regions of high clustering of foci. Column 1 is the MKDA statistic $m(v)$, column 2 is the ALE statistic $\ell(v)$, column 3 is the SDM statistic $s(v)$, and column 4 is the study activation center intensity $\sum_{i=1}^{I}{\left[\lambda_i^0(v)+\rho_i(v)\right]}$ for the BHICP. Rows 1-3 correspond to axial slices $z=-22$, $z=-16$ and $z=-2$, respectively. Note that differences in brain shapes across methods are simply due to different masks used by the various algorithms.}
      \label{fig:emotionresults}
 \end{sidewaysfigure}

Figure \ref{fig:emotionresults} shows statistic images obtained from the four methods presented above, conditional on several values of the $z$ dimension. 
Note that for ALE, MKDA and SDM we show $\ell(v)$, $m(v)$ and $s(v)$, respectively, whereas for the BHICP we show the voxel-wise posterior mean of the activation study intensity function, namely $\sum_{i=1}^{I}{\left[\lambda_i^0(v)+\rho_i(v)\right]}$\footnote{\citet{Kang2011} note that singleton foci can also be viewed as activation centers since they come directly from the population centers.}. 
We see that all of the methods provide qualitatively similar results. 
More specifically, the regions of the brain that are mostly engaged in emotion processing are the right and left amygdala (Fig.\ \ref{fig:emotionresults}, top and middle row). 
This finding is consistent with previous analyses of the same dataset \citep{Kober2008,Kang2011,Yue2012,Kang2014} as well as results of previous studies \citep{Phelps2005,Costafreda2008}. 
Other regions with moderately high values are the right and left cerebral cortex (Fig.\ \ref{fig:emotionresults}, bottom row). For the BHICP, this pattern is only noticeable in the right cerebral cortex.

\subsection{Discussion}\label{sec:reviewdiscussion}
In this Section, we build on our real data analysis to discuss the differences between kernel-based and model-based approaches in greater detail.

{\it Interpretation.} The meta-analysis of emotions suggests that the results obtained with model-based methods are qualitatively similar to those obtained with kernel-based methods. 
However, it is difficult to compare these results quantitatively as they have very different interpretations. 
At each voxel $v$, the MKDA statistic $m(v)$ represents the weighted proportion of studies that reported an activation within 10 mm of that given voxel. For the emotion dataset, the maximum value of $m(v)$, say $m(v^\ast)$, is $0.177$ and $v^\ast$ is located in the left amygdala. The
ALE statistic $\ell(v)$ estimates the probability that at least one of the foci closest to $v$ in each study is truly located at $v$. 
We observe a maximum of $0.158$ for $\ell(v)$ in our analysis. 
SDM uses ALE's normal kernel but combines study maps as MKDA, hence interpretation is difficult. 
However, $s(v)$ can be still viewed as a qualitative measure of foci clustering around $v$. 
The maximum value for SDM is 0.181, and $v^\ast$, the voxel where the maximum occurs, is also located near the left amygdala.     

In model-based methods, one obtains posterior distributions for all of the model parameters rather than a single statistic. Given the posterior samples, various estimates can be constructed. 
For the point process models of \cite{Kang2011}, \citet{Kang2014} and \citet{Montagna2016}, one mainly utilises the posterior draws of the intensity functions, say $\lambda^t(v)$, where $t$ indexes the MCMC samples. 
Conditional on $\lambda^t(v)$, the expected number of foci at any voxel $v$ has a Poisson distribution with mean $8\lambda^t(v)$, where 8 is the volume of each voxel on a standard $2\times 2\times2$mm mask. 
With the BHICP, one can infer the location of the activation centers or the location of the population centres. 
Instead, the PGRF and the latent factor model of \citet{Montagna2016} provide the expected number of foci for a given experimental paradigm. 
Further, the PGRF will produce one intensity per study-type $j$ whereas the latent factor model will produce one intensity map per study. 
Population results can be obtained by averaging the study-specific posterior intensity maps.      
Finally, the model of \cite{Yue2012} produces posterior draws $z^t(v)$ and $\psi^t(v)$, which can be used to estimate the probability that a voxel is activated, $H(z^t(v))$, and the probability that a voxel $v$ is falsely reported, $\mathbb{P}(\psi^t(v)=1)$.  
      
{\it Implementation \& computational considerations}. 
Software for kernel-based methods can be applied to any dataset and will produce a pair of brain images, one with the value of the statistic at each and every voxel and one containing the corresponding $p$-values. By directly comparing these two images, one can easily identify significant voxels. Instead, it is not straightforward to implement an MCMC scheme for the Bayesian model-based methods discussed in Section~\ref{sec:modelbased}. 
Prior specifications that are suitable for one dataset may be completely inappropriate for another. Further, it is not possible to know in advance how many iterations are required for the MCMC algorithm to converge, and convergence needs to be assessed as well. \\
\indent In terms of computing time, kernel-based methods outperform model-based methods (with the exception of SDM). 
For the emotion dataset in Section \ref{sec:reviewreal}, ALE required approximately 15 minutes of run time whereas MKDA required around 3 hours for 10,000 Monte Carlo replications. 
On the contrary, the BHICP model took roughly 16 hours for 120,000 MCMC iterations, \citet{Kang2014} needed 20 hours to complete the analysis, and it is yet not possible to run the spatial binary regression model on the full brain. 
\indent From a purely computational perspective, therefore, it may not seem appealing to a neuroimaging practitioner to adopt model-based methods, and this might explain why kernel-based methods are generally preferred. Kernel-based methods, however, suffer from some other serious limitations.  

\indent \textit{Explicit spatial modelling.} 
Kernel-based methods are based on a \textit{mass univariate approach} (MUA) that lacks an explicit spatial approach to the modelling of the foci. Specifically, MUA consists of fitting a model at each voxel independently of every other voxel. 
Even though the activation of nearby voxels is correlated, estimation with the MUA ignores the spatial correlation, but inference later accounts for it when random field theory or permutation procedures define a threshold for significant activation. 
By failing to capture the spatial nature of the data, kernel-based methods do not provide an accurate representation of the true data generating mechanism and can not jointly characterise randomness of the number and locations of activations within each study. 
With point process models instead, one can obtain the expected number of foci in any region of interest simply by integrating the intensity function over that region. 
For example, we can use the posterior draws of the BHICP model to infer that the expected number of activation centers in the amygdala is 2.1 (95\% CI [1.8,2.6]) for the emotion dataset, and that the median number of population centers in the entire brain is 7 (95\% CI [5,10]). 

Several other quantities of interest can be obtained from model-based methods. 
For the BHICP model, it is possible to derive $(1-\alpha)\%$ credible ellipses for both population and study activation centers, thus returning an estimate of within-study and between study-variability as in a random effects meta-analysis model. 
By introducing the latent process $\psi\left(v\right)$, \citet{Yue2012} estimate the probability of a voxel being miscoded. 
In the HPGRF model, the authors provide correlation estimates between the different emotions. 
All these quantities can not be obtained by any of the kernel-based methods.

By lacking an explicit spatial model, kernel-based methods can not deliver, for example, spatial confidence intervals and an arbitrary kernel size parameter ($r$ for MKDA and $\sigma$ for ALE and SDM) must be set (though \cite{Eickhoff2009} offer heuristics). 
Typically, its value is specified based on previous studies rather than being estimated from the data, and it remains constant across the brain regardless of the amount of smoothing required in each region. 
However, a bad choice for the kernel size can potentially affect the results.  
For example, in the third column of Figure \ref{fig:emotionresults} we observe bigger clusters resulting from a choice for a larger kernel size than that used for MKDA and ALE.  
\citet{Yue2012} automatically choose the required amount of smoothing by introducing an extra smoothness parameters in their GMRF.

\textit{Quantification of uncertainty.} There is no measure of uncertainty associated with the effect estimate in kernel-based methods, thus conclusions could be misleading. 
For example, in Section \ref{sec:reviewale} we found that power properties of ALE do not degrade with the inclusion of poor quality studies (see Fig.\ \ref{fig:alesim2}). 
Since inferences remain roughly unchanged, it is not possible to distinguish between cases with strong signal (few poor quality studies) and weak signal (many poor quality studies). 
Note that this is a fixed effects model property, where a small proportion of the data drives the inference. 
Model-based methods tackle this problem by providing standard errors obtained directly from the posterior distribution of any parameter of interest. When the signal is strong there will be small variability in the posterior estimates, whereas when signal is weak uncertainty will be higher.

\textit{Mechanisms for reverse inference.} The Bayesian framework upon which model-based methods are built facilitates the construction of predictive distributions over new studies. 
This helps producing the so-called {\em reverse inferences} \citep{Poldrack2011b}, a topic of growing interest in the fMRI community. 
Traditionally, fMRI studies produce {\em forward inferences}: for a given task or paradigm, inference is made on the location of the brain response to the task. 
Reverse inference consists of using the pattern of brain activation to infer which task is most likely to have produced the data. 
If a neuroscientist has developed a new behavioural experiment (say, on emotion), he/she may indeed want to know whether their task engages the brain's emotion processing system. 
In such case the researcher would want an estimate of the probability that the data arose from a population of emotion studies. 
\citet{Kang2014} show that classification based on the HPGRF model outperforms a naive classifier based on MKDA (MKDA + NBC). Reverse inference is naturally embedded into the spatial Bayesian latent factor model of \citet{Montagna2016}, and classification performance outperforms MKDA + NBC on a meta-analysis of emotion and executive control studies. These results suggest that spatial models can capture information in the data that can not be learnt with a MUA. 

To conclude, kernel-based methods can be more efficient than model-based methods when there is need for quick exploratory analyses, and for quick comparisons between several different datasets.  
Nevertheless, we believe that model-based methods have significant merits compared to kernel-based methods because of the extensive inference that can be performed when using these methodologies. 
Of course, there are still several open problems for model-based methods and CBMA in general. 
We discuss these open problems in the following section.

\section{Open problems}\label{sec:challenges}
Some aspects of CBMA are still being overlooked by both kernel-based and kernel-based methods. 
Arguably the most important is publication bias, which occurs when publicly accessible studies are not a representative subset of the total population of studies. A special case of publication bias is the so-called file drawer, that is, studies with significant findings are more likely to get published. 
If present, publication bias can affect the outcome of a meta-analysis and lead to false conclusions.  
While there is evidence for the existence of publication biases in fMRI \citep{David2013}, there has been no attempt to quantify neither the extent of the phenomenon nor the effect these biases may have on meta-analysis estimates. 

Another area of emerging interest is that of meta-regression, that is, the use of study-specific characteristics as explanatory variables in a meta-analysis model \citep{Greenland1994}. Meta-regression is an important facet of meta-analysis, especially when there is appreciable heterogeneity between studies. 
As noted by \citet{Carp2012}, fMRI experiments exhibit strong heterogeneity, hence it is essential to explore the effect that study characteristics have on the outcome of the analyses. 
Of both kernel and model-based approaches revised in Section \ref{sec:cbma}, only the spatial Bayesian latent factor regression model of \citet{Montagna2016} can  accommodate covariate information collected on the different studies. 
It is therefore a breakthrough contribution in neuroimaging meta-analysis research. 
By incorporating covariates in a meta-analysis on emotions, the authors observe, for example, that failing to adjust for multiple hypothesis testing results into a higher expected number of foci over the brain. 
The other CBMA methods described in Section \ref{sec:cbma} would assume that any deviance from the true population effect is due to sampling error. 
While current implementations of other model-based methods can not accommodate covariates, it is certainly possible to extend them to address this point. 
Thus, meta-regression remains an important topic that should be explored further in future research.

Another interesting avenue for future research consists in combining CBMA with image data from new fMRI studies. 
For example, there are no models that combine CBMA with IBMA when full 3D statistic images are available. Further, it is currently not possible to use meta-analysis data to improve estimation in small, underpowered group fMRI studies.  A plausible direction is to build a point process model for single studies where the prior distribution for the study centers corresponds to the posterior intensity as obtained from a point process CBMA and IBMA model.

Often, additional information is available about the foci, such as the corresponding $p$-values or $T$ scores. A convenient way to model these quantities is via mark processes, where these values become marks of the existing point patterns to improve estimation of the intensity function. Such an approach can enrich the inferences obtained from a meta-analysis by characterising the magnitude of activation at each voxel as opposed to modelling the location (and number) of the activations only, which is the question that current (model-based) methods address. 

Finally, there is little work on functional connectivity for model-based methods. 
Functional connectivity refers to the dependency between one or more regions of the brain. 
In CBMA functional connectivity is implied by co-activation, that is, when two regions consistently report activations.  
In a recent work, \citet{Xue2014} use a multivariate Poisson model to induce correlation among the foci in several regions on interest. 
However, it would be interesting to extend the spatial models in order to capture these correlations as well.

\section*{Acknowledgements}
The authors thank Jian Kang for helpful discussions and gratefully acknowledge the efforts of the labs of Tor Wager and Lisa Feldman Barrett for creating the emotion meta-analysis database and making it freely available. The authors also thank the two referees, the Associate Editor and the Editor for their constructive comments that helped improve the manuscript substantially. This work was largely completed while PS, SM and TEN were at the University of Warwick, Department of Statistics. PS, TEN \& TDJ were supported by NIH grant 5-R01-NS-075066; TEN was supported by a Wellcome Trust fellowship 100309/Z/12/Z and NIH grant R01 2R01EB015611-04. The work presented in this paper represents the views of the authors and not necessarily those of the NIH or the Wellcome Trust Foundation.

\bibliography{bibo.bib}

\begin{thebibliography}{}

\bibitem[Arminger and Muth\'{e}n(1998)Arminger and Muth\'{e}n]{Arminger1998}
Arminger, G. and Muth\'{e}n, B.~O. (1998).
\newblock {A Bayesian approach to nonlinear latent variable models using the
  Gibbs sampler and the Metropolis-Hastings algorithm}.
\newblock {\em Psychometrika\/}, {\bf 63}(3), 271--300.

\bibitem[Bailey {\em et~al.}(2006)Bailey, Townsend, Valk, and
  Maisey]{Bailey2006}
Bailey, D., Townsend, D., Valk, P., and Maisey, M., editors (2006).
\newblock {\em Positron Emission Tomography: Basic Sciences\/}.
\newblock Springer-Verlag.

\bibitem[Bartels and Zeki(2004)Bartels and Zeki]{Bartels2004}
Bartels, A. and Zeki, S. (2004).
\newblock {The neural correlates of maternal and romantic love}.
\newblock {\em Neuroimage\/}, {\bf 21}(3), 1155--1166.

\bibitem[Benjamini and Hochberg(1995)Benjamini and Hochberg]{Benjamini1995}
Benjamini, Y. and Hochberg, Y. (1995).
\newblock Controlling the false discovery rate: A practical and powerful
  approach to multiple testing.
\newblock {\em Journal of the Royal Statistical Society. Series B
  (Methodological)\/}, {\bf 57}(1), pp. 289--300.

\bibitem[Button {\em et~al.}(2013)Button, Ioannidis, Mokrysz, Nosek, Flint,
  Robinson, and Munaf\`{o}]{Button2013}
Button, K.~S., Ioannidis, J. P.~a., Mokrysz, C., Nosek, B.~a., Flint, J.,
  Robinson, E. S.~J., and Munaf\`{o}, M.~R. (2013).
\newblock {Power failure: why small sample size undermines the reliability of
  neuroscience.}
\newblock {\em Nature reviews. Neuroscience\/}, {\bf 14}(5), 365--76.

\bibitem[Calhoun and Pearlson(2012)Calhoun and Pearlson]{Calhoun2012}
Calhoun, V.~D. and Pearlson, G.~D. (2012).
\newblock {A selective review of simulated driving studies: Combining
  naturalistic and hybrid paradigms, analysis approaches, and future
  directions.}
\newblock {\em NeuroImage\/}, {\bf 59}(1), 25--35.

\bibitem[Carp(2012)Carp]{Carp2012}
Carp, J. (2012).
\newblock The secret lives of experiments: Methods reporting in the fmri
  literature.
\newblock {\em NeuroImage\/}, {\bf 63}(1), 289 -- 300.

\bibitem[Caspers {\em et~al.}(2014)Caspers, Zilles, Beierle, Rottschy, and
  Eickhoff]{Caspers2014}
Caspers, J., Zilles, K., Beierle, C., Rottschy, C., and Eickhoff, S.~B. (2014).
\newblock A novel meta-analytic approach: Mining frequent co-activation
  patterns in neuroimaging databases.
\newblock {\em NeuroImage\/}, {\bf 90}(0), 390 -- 402.

\bibitem[Cole {\em et~al.}(2010)Cole, Smith, and Beckmann]{Cole2010}
Cole, D.~M., Smith, S.~M., and Beckmann, C.~F. (2010).
\newblock {Advances and pitfalls in the analysis and interpretation of
  resting-state fMRI data.}
\newblock {\em Frontiers in systems neuroscience\/}, {\bf 4}(8).

\bibitem[Costafreda {\em et~al.}(2008)Costafreda, Brammer, David, and
  Fu]{Costafreda2008}
Costafreda, S., Brammer, M., David, A., and Fu, C. (2008).
\newblock Predictors of amygdala activation during the processing of emotional
  stimuli.
\newblock {\em Brain Research Reviews\/}, {\bf 58}(1), 57 -- 70.

\bibitem[Costafreda {\em et~al.}(2009)Costafreda, David, and
  Brammer]{Costafreda2009}
Costafreda, S.~G., David, A.~S., and Brammer, M.~J. (2009).
\newblock A parametric approach to voxel-based meta-analysis.
\newblock {\em NeuroImage\/}, {\bf 46}(1), 115 -- 122.

\bibitem[David {\em et~al.}(2013)David, Ware, Chu, Loftus, Fusar-Poli, Radua,
  Munaf\`{o}, and Ioannidis]{David2013}
David, S.~P., Ware, J.~J., Chu, I.~M., Loftus, P.~D., Fusar-Poli, P., Radua,
  J., Munaf\`{o}, M.~R., and Ioannidis, J. P.~A. (2013).
\newblock {Potential Reporting Bias in fMRI Studies of the Brain}.
\newblock {\em PLoS ONE\/}, {\bf 8}(7).

\bibitem[Delvecchio {\em et~al.}(2012)Delvecchio, Fossati, Boyer, Brambilla,
  Falkai, Gruber, Hietala, Lawrie, Martinot, McIntosh, Meisenzahl, and
  Frangou]{Delvecchio2012}
Delvecchio, G., Fossati, P., Boyer, P., Brambilla, P., Falkai, P., Gruber, O.,
  Hietala, J., Lawrie, S.~M., Martinot, J.-L., McIntosh, A.~M., Meisenzahl, E.,
  and Frangou, S. (2012).
\newblock {Common and distinct neural correlates of emotional processing in
  Bipolar Disorder and Major Depressive Disorder: a voxel-based meta-analysis
  of functional magnetic resonance imaging studies.}
\newblock {\em European neuropsychopharmacology : the journal of the European
  College of Neuropsychopharmacology\/}, {\bf 22}(2), 100--13.

\bibitem[Eickhoff {\em et~al.}(2009)Eickhoff, Laird, Grefkes, Wang, Zilles, and
  Fox]{Eickhoff2009}
Eickhoff, S.~B., Laird, A.~R., Grefkes, C., Wang, L.~E., Zilles, K., and Fox,
  P.~T. (2009).
\newblock Coordinate-based activation likelihood estimation meta-analysis of
  neuroimaging data: A random-effects approach based on empirical estimates of
  spatial uncertainty.
\newblock {\em Human Brain Mapping\/}, {\bf 30}(9), 2907--2926.

\bibitem[Eickhoff {\em et~al.}(2012)Eickhoff, Bzdok, Laird, Kurth, and
  Fox]{Eickhoff2012}
Eickhoff, S.~B., Bzdok, D., Laird, A.~R., Kurth, F., and Fox, P.~T. (2012).
\newblock Activation likelihood estimation meta-analysis revisited.
\newblock {\em NeuroImage\/}, {\bf 59}(3), 2349 -- 2361.

\bibitem[Eickhoff {\em et~al.}(2016)Eickhoff, Nichols, Laird, Hoffstaedter,
  Amunts, Fox, Bzdok, and Eickhoff]{Eickhoff2016}
Eickhoff, S.~B., Nichols, T.~E., Laird, A.~R., Hoffstaedter, F., Amunts, K.,
  Fox, P.~T., Bzdok, D., and Eickhoff, C.~R. (2016).
\newblock Behavior, sensitivity, and power of activation likelihood estimation
  characterized by massive empirical simulation.
\newblock {\em NeuroImage\/}, {\bf 137}, 70 -- 85.

\bibitem[Eickhoff {\em et~al.}(2017)Eickhoff, Laird, Fox, Lancaster, and
  Fox]{Eickhoff2017}
Eickhoff, S.~B., Laird, A.~R., Fox, P.~M., Lancaster, J.~L., and Fox, P.~T.
  (2017).
\newblock Implementation errors in the gingerale software: Description and
  recommendations.
\newblock {\em Human Brain Mapping\/}, {\bf 38}(1), 7--11.

\bibitem[Etkin and Wager(2007)Etkin and Wager]{Etkin2007}
Etkin, A. and Wager, T. (2007).
\newblock {Functional Neuroimaging of Anxiety : A Meta-Analysis of Emotional
  Processing in PTSD , Social Anxiety Disorder , and Specific Phobia}.
\newblock {\em American Journal of Psychiatry\/}, {\bf 164}(10), 1476--1488.

\bibitem[Farah(2014)Farah]{Farah2014}
Farah, M.~J. (2014).
\newblock {Brain images, babies, and bathwater: critiquing critiques of
  functional neuroimaging.}
\newblock {\em The Hastings Center report\/}, {\bf 44}(S2), S19--30.

\bibitem[Fox {\em et~al.}(1997)Fox, Lancaster, Parsons, Xiong, and
  Zamarripa]{Fox1997}
Fox, P.~T., Lancaster, J.~L., Parsons, L.~M., Xiong, J.~H., and Zamarripa, F.
  (1997).
\newblock {Functional volumes modeling: Theory and preliminary assessment}.
\newblock {\em Human Brain Mapping\/}, {\bf 5}(4), 306--311.

\bibitem[Fox {\em et~al.}(1998)Fox, Parsons, and Lancaster]{Fox1998}
Fox, P.~T., Parsons, L.~M., and Lancaster, J.~L. (1998).
\newblock {Beyond the single study: function/location metanalysis in cognitive
  neuroimaging}.
\newblock {\em Current Opinion in Neurobiology\/}, {\bf 8}(2), 178--187.

\bibitem[Friston {\em et~al.}(2002)Friston, Penny, Phillips, Kiebel, Hinton,
  and Ashburner]{Friston2002}
Friston, K.~J., Penny, W., Phillips, C., Kiebel, S., Hinton, G., and Ashburner,
  J. (2002).
\newblock Classical and bayesian inference in neuroimaging: Theory.
\newblock {\em NeuroImage\/}, {\bf 16}(2), 465 -- 483.

\bibitem[Fusar-Poli(2012)Fusar-Poli]{FusarPoli2012}
Fusar-Poli, P. (2012).
\newblock {Voxel-wise meta-analysis of fMRI studies in patients at clinical
  high risk for psychosis.}
\newblock {\em Journal of psychiatry \& neuroscience : JPN\/}, {\bf 37}(2),
  106--12.

\bibitem[Genovese {\em et~al.}(2002)Genovese, Lazar, and Nichols]{Genovese2002}
Genovese, C., Lazar, N., and Nichols, T. (2002).
\newblock {Thresholding of statistical maps in functional neuroimaging using
  the false discovery rate.}
\newblock {\em NeuroImage\/}, {\bf 15}(4), 870--878.

\bibitem[Gorgolewski {\em et~al.}(2015)Gorgolewski, Varoquaux, Rivera, Schwarz,
  Ghosh, Maumet, Sochat, Nichols, Poldrack, Poline, Yarkoni, and
  Margulies]{Gorgo2015}
Gorgolewski, K.~J., Varoquaux, G., Rivera, G., Schwarz, Y., Ghosh, S.~S.,
  Maumet, C., Sochat, V.~V., Nichols, T.~E., Poldrack, R.~A., Poline, J.-B.,
  Yarkoni, T., and Margulies, D.~S. (2015).
\newblock Neurovault.org: a web-based repository for collecting and sharing
  unthresholded statistical maps of the human brain.
\newblock {\em Frontiers in Neuroinformatics\/}, {\bf 9}, 8.

\bibitem[Greenland(1994)Greenland]{Greenland1994}
Greenland, S. (1994).
\newblock Invited commentary: A critical look at some popular meta-analytic
  methods.
\newblock {\em American Journal of Epidemiology\/}, {\bf 140}(3), 290--296.

\bibitem[Hartung {\em et~al.}(2008)Hartung, Knapp, and Sinha]{Hartung2008}
Hartung, J., Knapp, G., and Sinha, B.~K. (2008).
\newblock {\em Statistical Meta-Analysis with Applications\/}.
\newblock John Wiley \& Sons, Hoboken.

\bibitem[Hedges and Olkin(1985)Hedges and Olkin]{Hedges1985}
Hedges, L.~V. and Olkin, I. (1985).
\newblock {\em Statistical Methods for Meta-analysis\/}.
\newblock Academic Press.

\bibitem[Huettel {\em et~al.}(2009)Huettel, Song, and McCarthy]{Huettel2009}
Huettel, S.~A., Song, A.~W., and McCarthy, G. (2009).
\newblock {\em Functional Magnetic Resonance Imaging Second Edition\/}.
\newblock Sinauer Associates, Inc, Massachussets.

\bibitem[Illian {\em et~al.}(2008)Illian, Penttinen, Stoyan, and
  Stoyan]{Illian2008}
Illian, J., Penttinen, P., Stoyan, H., and Stoyan, D. (2008).
\newblock {\em Statistical Analysis and Modelling of Spatial Point Patterns\/}.
\newblock Wiley.

\bibitem[Kang {\em et~al.}(2011)Kang, Johnson, Nichols, and Wager]{Kang2011}
Kang, J., Johnson, T.~D., Nichols, T.~E., and Wager, T.~D. (2011).
\newblock {Meta Analysis of Functional Neuroimaging Data via Bayesian Spatial
  Point Processes}.
\newblock {\em Journal of the American Statistical Association\/}, {\bf
  106}(493), 124--134.

\bibitem[Kang {\em et~al.}(2014)Kang, Nichols, Wager, and Johnson]{Kang2014}
Kang, J., Nichols, T.~E., Wager, T.~D., and Johnson, T.~D. (2014).
\newblock {A bayesian hierarchical spatial point process model for multi-type
  neuroimaging meta-analysis}.
\newblock {\em The Annals of Applied Statistics\/}, {\bf 8}(3), 1561--1582.

\bibitem[Kim and Ogawa(2012)Kim and Ogawa]{Kim2012}
Kim, S.-G. and Ogawa, S. (2012).
\newblock {Biophysical and physiological origins of blood oxygenation
  level-dependent fMRI signals.}
\newblock {\em Journal of cerebral blood flow and metabolism : official journal
  of the International Society of Cerebral Blood Flow and Metabolism\/}, {\bf
  32}(7), 1188--206.

\bibitem[Kober {\em et~al.}(2008)Kober, Barrett, Joseph, Bliss-Moreau,
  Lindquist, and Wager]{Kober2008}
Kober, H., Barrett, L.~F., Joseph, J., Bliss-Moreau, E., Lindquist, K., and
  Wager, T.~D. (2008).
\newblock Functional grouping and cortical and subcortical interactions in
  emotion: A meta-analysis of neuroimaging studies.
\newblock {\em NeuroImage\/}, {\bf 42}(2), 998 -- 1031.

\bibitem[Konova {\em et~al.}(2013)Konova, Moeller, and Goldstein]{Konova2013}
Konova, A.~B., Moeller, S.~J., and Goldstein, R.~Z. (2013).
\newblock Common and distinct neural targets of treatment: Changing brain
  function in substance addiction.
\newblock {\em Neuroscience \& Biobehavioral Reviews\/}, {\bf 37}(10),
  2806--2817.

\bibitem[Laird {\em et~al.}(2005a)Laird, Fox, Price, Glahn, Uecker, Lancaster,
  Turkeltaub, Kochunov, and Fox]{Laird2005b}
Laird, A.~R., Fox, P.~M., Price, C.~J., Glahn, D.~C., Uecker, A.~M., Lancaster,
  J.~L., Turkeltaub, P.~E., Kochunov, P., and Fox, P.~T. (2005a).
\newblock Ale meta-analysis: Controlling the false discovery rate and
  performing statistical contrasts.
\newblock {\em Human Brain Mapping\/}, {\bf 25}(1), 155--164.

\bibitem[Laird {\em et~al.}(2005b)Laird, Lancaster, and Fox]{Laird2005a}
Laird, A.~R., Lancaster, J.~J., and Fox, P.~T. (2005b).
\newblock Brainmap: The social evolution of a human brain mapping database.
\newblock {\em Neuroinformatics\/}, {\bf 3}(1), 65--77.

\bibitem[Lazar {\em et~al.}(2002)Lazar, Luna, Sweeney, and Eddy]{Lazar2002}
Lazar, N.~A., Luna, B., Sweeney, J.~A., and Eddy, W.~F. (2002).
\newblock Combining brains: A survey of methods for statistical pooling of
  information.
\newblock {\em NeuroImage\/}, {\bf 16}(2), 538 -- 550.

\bibitem[Lindquist(2008)Lindquist]{Lindquist2008}
Lindquist, M.~A. (2008).
\newblock The statistical analysis of f{MRI} data.
\newblock {\em Statistical Science\/}, {\bf 23}(4), 439--464.

\bibitem[M{\o}ller and Waagepetersen(2004)M{\o}ller and
  Waagepetersen]{Moller2004}
M{\o}ller, J. and Waagepetersen, R.~P. (2004).
\newblock {\em Statistical Inference and Simulation for Spatial Point
  Processes\/}.
\newblock Chapman and Hall/CRC, Boca Raton.

\bibitem[Montagna {\em et~al.}(2012)Montagna, Tokdar, Neelon, and
  Dunson]{Montagna2012}
Montagna, S., Tokdar, S.~T., Neelon, B., and Dunson, D.~B. (2012).
\newblock {Bayesian latent factor regression for functional and longitudinal
  data.}
\newblock {\em Biometrics\/}, {\bf 68}(4), 1064--73.

\bibitem[{Montagna} {\em et~al.}(2017){Montagna}, {Wager}, {Feldman-Barrett},
  {Johnson}, and {Nichols}]{Montagna2016}
{Montagna}, S., {Wager}, T., {Feldman-Barrett}, L., {Johnson}, T.~D., and
  {Nichols}, T.~E. (2017).
\newblock Spatial bayesian latent factor regression modeling of
  coordinate-based meta-analysis data.
\newblock (In press).

\bibitem[Mumford and Nichols(2006)Mumford and Nichols]{Mumford2006}
Mumford, J.~A. and Nichols, T. (2006).
\newblock Modeling and inference of multisubject fmri data.
\newblock {\em Engineering in Medicine and Biology Magazine, IEEE\/}, {\bf
  25}(2), 42 --51.

\bibitem[Mumford and Nichols(2009)Mumford and Nichols]{Mumford2009}
Mumford, J.~A. and Nichols, T. (2009).
\newblock Simple group fmri modeling and inference.
\newblock {\em NeuroImage\/}, {\bf 47}(4), 1469 -- 1475.

\bibitem[Nichols and Hayasaka(2003)Nichols and Hayasaka]{Hayasaka2003}
Nichols, T. and Hayasaka, S. (2003).
\newblock Controlling the familywise error rate in functional neuroimaging: a
  comparative review.
\newblock {\em Statistical Methods in Medical Research\/}, {\bf 12}(5),
  419--446.

\bibitem[Nichols and Holmes(2002)Nichols and Holmes]{Nichols2002}
Nichols, T.~E. and Holmes, A.~P. (2002).
\newblock {Nonparametric permutation tests for functional neuroimaging: a
  primer with examples.}
\newblock {\em Human brain mapping\/}, {\bf 15}(1), 1--25.

\bibitem[Nielsen and Hansen(2002)Nielsen and Hansen]{Nielsen2002}
Nielsen, F.~{\AA}. and Hansen, L.~K. (2002).
\newblock {Modeling of activation data in the BrainMap™ database: Detection
  of outliers}.
\newblock {\em Human Brain Mapping\/}, {\bf 15}(3), 146--156.

\bibitem[Phelps and LeDoux(2005)Phelps and LeDoux]{Phelps2005}
Phelps, E.~A. and LeDoux, J.~E. (2005).
\newblock {Contributions of the amygdala to emotion processing: From animal
  models to human behavior}.
\newblock {\em Neuron\/}, {\bf 48}(2), 175--187.

\bibitem[Poldrack(2011)Poldrack]{Poldrack2011b}
Poldrack, R. (2011).
\newblock {Inferring Mental States from Neuroimaging Data: From Reverse
  Inference to Large-Scale Decoding}.

\bibitem[Poldrack {\em et~al.}(2011)Poldrack, Mumford, and
  Nichols]{Poldrack2011}
Poldrack, R., Mumford, J., and Nichols, T. (2011).
\newblock {\em Handbook of Functional MRI Data Analysis\/}.
\newblock Cambridge University Press, Cambridge.

\bibitem[Poldrack {\em et~al.}(2013)Poldrack, Barch, Mitchell, Wager, Wagner,
  Devlin, Cumba, Koyejo, and Milham]{Poldrack2013}
Poldrack, R., Barch, D., Mitchell, J., Wager, T., Wagner, A., Devlin, J.,
  Cumba, C., Koyejo, O., and Milham, M. (2013).
\newblock Toward open sharing of task-based fmri data: the openfmri project.
\newblock {\em Frontiers in Neuroinformatics\/}, {\bf 7}, 12.

\bibitem[Poline {\em et~al.}(2012)Poline, Breeze, Ghosh, Gorgolewski,
  Halchenko, Hanke, Helmer, Marcus, Poldrack, Schwartz, Ashburner, and
  Kennedy]{Poline2012}
Poline, J.-B., Breeze, J., Ghosh, S., Gorgolewski, K., Halchenko, Y., Hanke,
  M., Helmer, K., Marcus, D., Poldrack, R., Schwartz, Y., Ashburner, J., and
  Kennedy, D. (2012).
\newblock Data sharing in neuroimaging research.
\newblock {\em Frontiers in Neuroinformatics\/}, {\bf 6}, 9.

\bibitem[Radua and Mataix-Cols(2009)Radua and Mataix-Cols]{Radua2009}
Radua, J. and Mataix-Cols, D. (2009).
\newblock {Voxel-wise meta-analysis of grey matter changes in
  obsessive-compulsive disorder.}
\newblock {\em The British journal of psychiatry : the journal of mental
  science\/}, {\bf 195}(5), 393--402.

\bibitem[Radua and Mataix-Cols(2012)Radua and Mataix-Cols]{Radua2012b}
Radua, J. and Mataix-Cols, D. (2012).
\newblock {Meta-analytic methods for neuroimaging data explained.}
\newblock {\em Biology of mood \& anxiety disorders\/}, {\bf 2}(1), 6.

\bibitem[Radua {\em et~al.}(2012)Radua, Mataix-Cols, Phillips, El-Hage,
  Kronhaus, Cardoner, and Surguladze]{Radua2012}
Radua, J., Mataix-Cols, D., Phillips, M.~L., El-Hage, W., Kronhaus, D.~M.,
  Cardoner, N., and Surguladze, S. (2012).
\newblock {A new meta-analytic method for neuroimaging studies that combines
  reported peak coordinates and statistical parametric maps}.
\newblock {\em European Psychiatry\/}, {\bf 27}(8), 605--611.

\bibitem[Radua {\em et~al.}(2014)Radua, Rubia, Canales-Rodr\'{\i}guez,
  Pomarol-Clotet, Fusar-Poli, and Mataix-Cols]{Radua2014}
Radua, J., Rubia, K., Canales-Rodr\'{\i}guez, E.~J., Pomarol-Clotet, E.,
  Fusar-Poli, P., and Mataix-Cols, D. (2014).
\newblock {Anisotropic kernels for coordinate-based meta-analyses of
  neuroimaging studies.}
\newblock {\em Frontiers in psychiatry\/}, {\bf 5}(February), 13.

\bibitem[Raemaekers {\em et~al.}(2007)Raemaekers, Vink, Zandbelt, van Wezel,
  Kahn, and Ramsey]{Raemaekers2007}
Raemaekers, M., Vink, M., Zandbelt, B., van Wezel, R., Kahn, R., and Ramsey, N.
  (2007).
\newblock Test-retest reliability of fmri activation during prosaccades and
  antisaccades.
\newblock {\em NeuroImage\/}, {\bf 36}(3), 532 -- 542.

\bibitem[Richlan {\em et~al.}(2011)Richlan, Kronbichler, and
  Wimmer]{Richlan2011}
Richlan, F., Kronbichler, M., and Wimmer, H. (2011).
\newblock {Meta-analyzing brain dysfunctions in dyslexic children and adults.}
\newblock {\em NeuroImage\/}, {\bf 56}(3), 1735--42.

\bibitem[Salimi-Khorshidi {\em et~al.}(2009)Salimi-Khorshidi, Smith, Keltner,
  Wager, and Nichols]{Salimi2009}
Salimi-Khorshidi, G., Smith, S.~M., Keltner, J.~R., Wager, T.~D., and Nichols,
  T.~E. (2009).
\newblock Meta-analysis of neuroimaging data: A comparison of image-based and
  coordinate-based pooling of studies.
\newblock {\em NeuroImage\/}, {\bf 45}(3), 810 -- 823.

\bibitem[Shermer(2008)Shermer]{Shermer2008}
Shermer, M. (2008).
\newblock {Why You Should Be Skeptical of Brain Scans}.
\newblock {\em Scientific American Mind\/}, {\bf 19}(5), 66--71.

\bibitem[Talairach and Tournoux(1988)Talairach and Tournoux]{Talairach1988}
Talairach, J. and Tournoux, P. (1988).
\newblock {\em Co-planar Stereotaxic Atlas of the Human Brain\/}.
\newblock Thieme, Stuttgart.

\bibitem[Thirion {\em et~al.}(2007)Thirion, Pinel, Meriaux, Roche, Dehaene, and
  Poline]{Thirion2007}
Thirion, B., Pinel, P., Meriaux, S., Roche, A., Dehaene, S., and Poline, J.-B.
  (2007).
\newblock Analysis of a large f{MRI} cohort: Statistical and methodological
  issues for group analyses.
\newblock {\em NeuroImage\/}, {\bf 35}(1), 105 -- 120.

\bibitem[Turkeltaub {\em et~al.}(2002)Turkeltaub, Eden, Jones, and
  Zeffiro]{Turkeltaub2001}
Turkeltaub, P.~E., Eden, G.~F., Jones, K.~M., and Zeffiro, T.~A. (2002).
\newblock Meta-analysis of the functional neuroanatomy of single-word reading:
  Method and validation.
\newblock {\em NeuroImage\/}, {\bf 16}(3, Part A), 765 -- 780.

\bibitem[Turkeltaub {\em et~al.}(2012)Turkeltaub, Eickhoff, Laird, Fox, Wiener,
  and Fox]{Turkeltaub2012}
Turkeltaub, P.~E., Eickhoff, S.~B., Laird, A.~R., Fox, M., Wiener, M., and Fox,
  P. (2012).
\newblock Minimizing within-experiment and within-group effects in activation
  likelihood estimation meta-analyses.
\newblock {\em Human Brain Mapping\/}, {\bf 33}(1), 1--13.

\bibitem[Vul {\em et~al.}(2009)Vul, Harris, Winkielman, and Pashler]{Vul2009c}
Vul, E., Harris, C., Winkielman, P., and Pashler, H. (2009).
\newblock {Puzzlingly High Correlations in fMRI Studies of Emotion,
  Personality, and Social Cognition}.
\newblock {\em Perspectives on Psychological Science\/}, {\bf 4}(3), 274--290.

\bibitem[Wager {\em et~al.}(2007)Wager, Lindquist, and Kaplan]{Wager2007}
Wager, T., Lindquist, M., and Kaplan, L. (2007).
\newblock Meta-analysis of functional neuroimaging data: current and future
  directions.
\newblock {\em Social Cognitive and Affective Neuroscience\/}, {\bf 2}(2),
  150--158.

\bibitem[Wager {\em et~al.}(2003)Wager, Phan, Liberzon, and Taylor]{Wager2003}
Wager, T.~D., Phan, K., Liberzon, I., and Taylor, S.~F. (2003).
\newblock {Valence, gender, and lateralization of functional brain anatomy in
  emotion: a meta-analysis of findings from neuroimaging}.
\newblock {\em NeuroImage\/}, {\bf 19}(3), 513--531.

\bibitem[Wager {\em et~al.}(2004)Wager, Jonides, and Reading]{Wager2004}
Wager, T.~D., Jonides, J., and Reading, S. (2004).
\newblock Neuroimaging studies of shifting attention: a meta-analysis.
\newblock {\em NeuroImage\/}, {\bf 22}(4), 1679 -- 1693.

\bibitem[Wager {\em et~al.}(2008)Wager, Barrett, Bliss-Moreau, Lindquist,
  Duncan, Kober, Joseph, Davidson, and Mize]{Wager2008}
Wager, T.~D., Barrett, L.~F., Bliss-Moreau, E., Lindquist, K., Duncan, S.,
  Kober, H., Joseph, J., Davidson, M., and Mize, J. (2008).
\newblock The neuroimaging of emotion.
\newblock In M.~Lewis, J.~Haviland-Jones, and L.~Barrett, editors, {\em
  Handbook of emotions\/}, pages 249--271. Guilford Press New York, NY, 3
  edition.

\bibitem[Wolpert and Ickstadt(1998)Wolpert and Ickstadt]{Wolpert1998}
Wolpert, R.~L. and Ickstadt, K. (1998).
\newblock Poisson/gamma random field models for spatial statistics.
\newblock {\em Biometrika\/}, {\bf 85}(2), 251--267.

\bibitem[Xue {\em et~al.}(2014)Xue, Kang, Bowman, Wager, and Guo]{Xue2014}
Xue, W., Kang, J., Bowman, F.~D., Wager, T.~D., and Guo, J. (2014).
\newblock Identifying functional co-activation patterns in neuroimaging studies
  via poisson graphical models.
\newblock {\em Biometrics\/}, {\bf 70}(4), 812--822.

\bibitem[Yarkoni {\em et~al.}(2011)Yarkoni, Poldrack, Nichols, Van~Essen, and
  Wager]{Yarkoni2011}
Yarkoni, T., Poldrack, R.~A., Nichols, T.~E., Van~Essen, D.~C., and Wager,
  T.~D. (2011).
\newblock Large-scale automated synthesis of human functional neuroimaging
  data.
\newblock {\em Nature methods\/}, {\bf 8}(8), 665--670.

\bibitem[Yue and Speckman(2010)Yue and Speckman]{Yue2010}
Yue, Y. and Speckman, P.~L. (2010).
\newblock Nonstationary spatial gaussian markov random fields.
\newblock {\em Journal of Computational and Graphical Statistics\/}, {\bf
  19}(1), 96--116.

\bibitem[Yue {\em et~al.}(2012)Yue, Lindquist, and Loh]{Yue2012}
Yue, Y.~R., Lindquist, M.~A., and Loh, J.~M. (2012).
\newblock {Meta-analysis of functional neuroimaging data using Bayesian
  nonparametric binary regression}.
\newblock {\em The Annals of Applied Statistics\/}, {\bf 6}(2), 697--718.

\end{thebibliography}

\newpage
\appendix

\section{Supplementary simulation results}\label{sec:app2}
In this Section, we repeat the simulations of Section \ref{sec:reviewale} using both MKDA and SDM kernels instead of the ALE kernel. 
The objective of this study is to assess the robustness of the results of Section \ref{sec:reviewale} to the choice of kernel. 
To reduce computation time, we use ALE's Monte Carlo significance test \citep{Eickhoff2012} for inference instead of the MKDA and SDM tests. 
For MKDA, the kernel size is set to $r=10\mathrm{mm}$, the default value of the MKDA software. 
For SDM, we set the standard deviation of the normal kernel to 4mm, the same used for ALE. 
Finally, we use the same synthetic datasets and consider power properties $1-4$ described in Section \ref{sec:reviewale}.

\begin{sidewaysfigure}[h]
        \centering
        \includegraphics[scale=1.8]{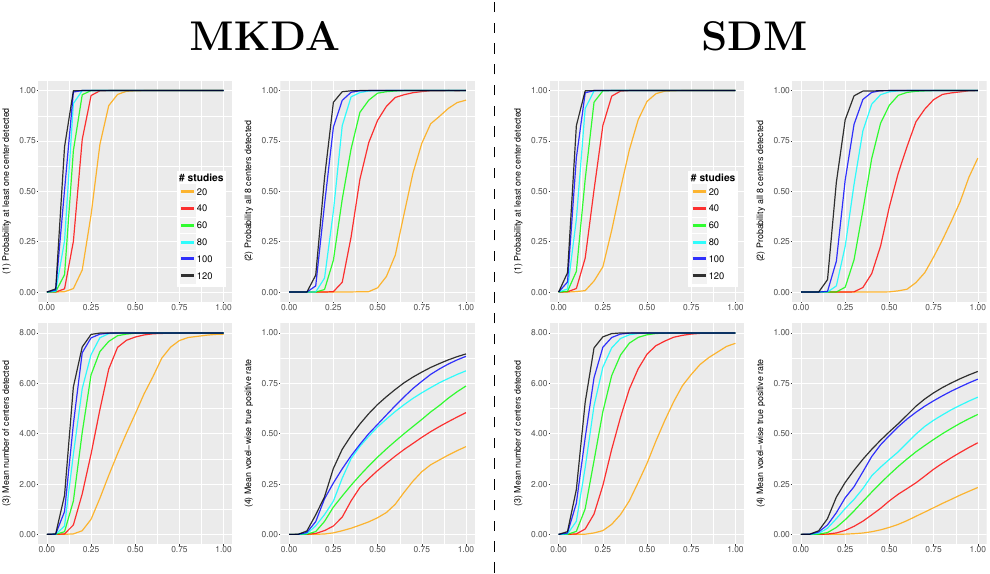}\\
        \vspace{-0.0in}
\caption{Power properties against the proportion of valid studies $p$, using the MKDA (left panel) and SDM (right panel) kernels. Top left (both panels): probability that at least one center is detected. Top right (both panels): probability that all 8 centers are detected. Bottom left (both panels): mean number of centers detected. Bottom right (both panels): mean voxel-wise true positive rate.}
      \label{fig:kernels1}
 \end{sidewaysfigure}

\begin{sidewaysfigure}[h]
        \centering
        \includegraphics[scale=1.8]{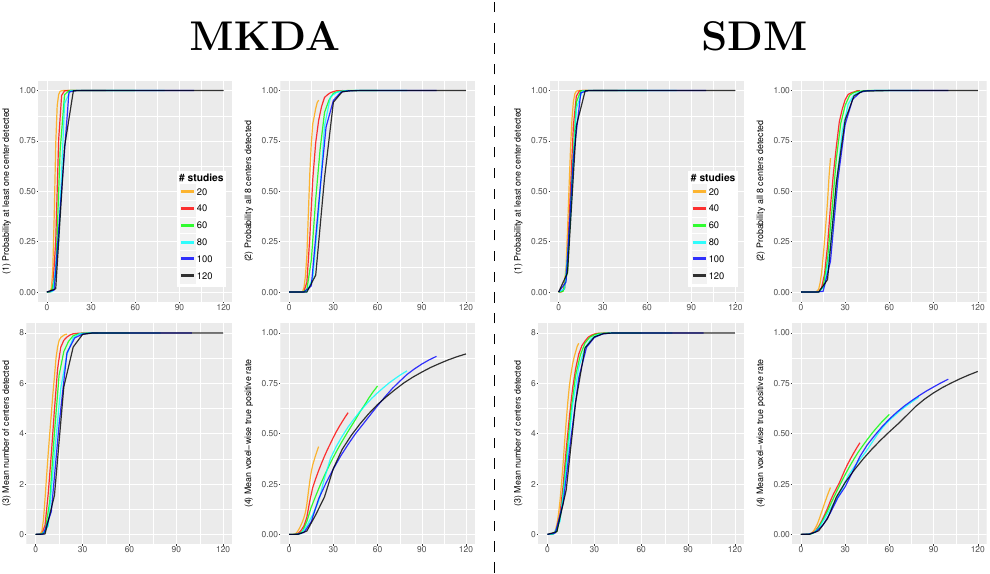}\\
        \vspace{-0.0in}
\caption{Power properties against the total number of valid studies $Ip$, using the MKDA (left panel) and SDM (right panel) kernels. Top left (both panels): probability that at least one center is detected. Top right (both panels): probability that all 8 centers are detected. Bottom left (both panels): mean number of centers detected. Bottom right (both panels): mean voxel-wise true positive rate.}
      \label{fig:kernels2}
 \end{sidewaysfigure}

Results are summarised in Figures \ref{fig:kernels1} and \ref{fig:kernels2}, where we present power properties $1-4$ as a function of the proportion and the absolute number of valid studies, respectively. 
We see that the consistency and robustness to noise properties of the ALE algorithm are maintained despite the use of a different kernel.

\section{MCMC convergence diagnostics for the BHICP model}\label{sec:app1}
\begin{sidewaysfigure}[h]
        \centering
        \includegraphics[scale=1.4]{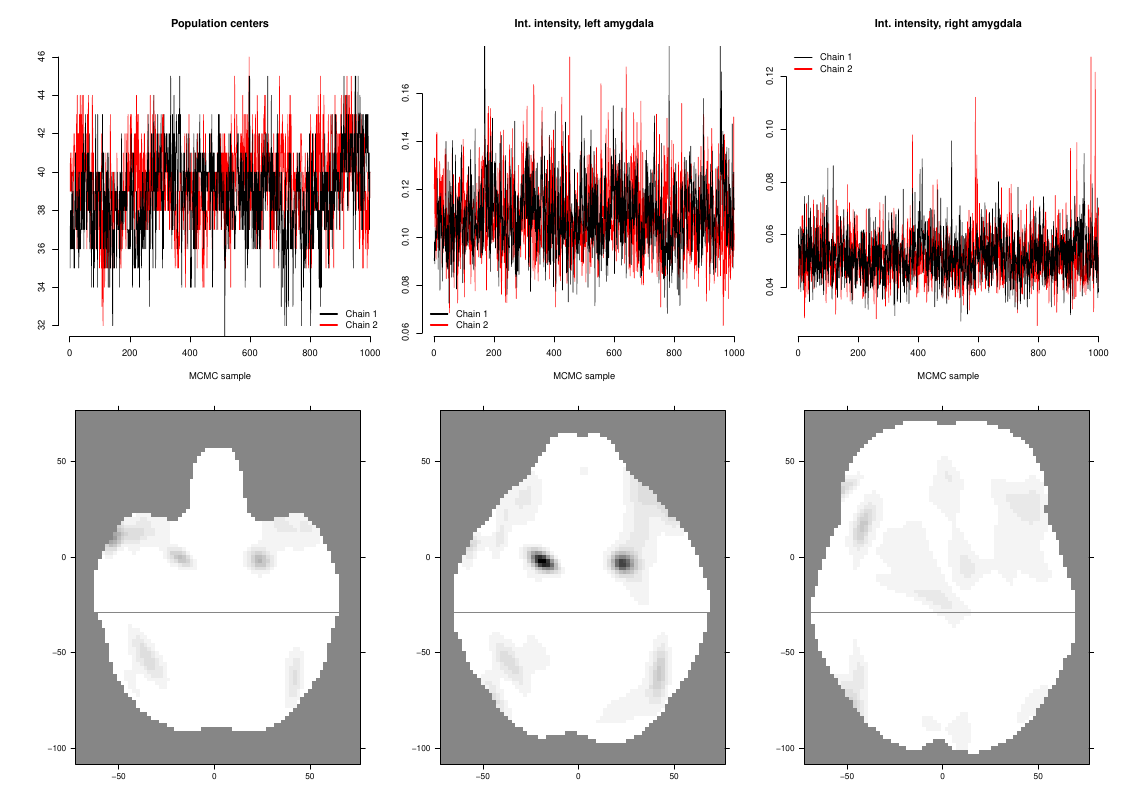}\\
        \vspace{-0.0in}
\caption{Convergence diagnostics for the BHICP analysis of the emotions dataset. Top panel: traceplots for the total number of population centers (left), integrated intensity in the left amygdala (middle), and integrated intensity in the right amygdala (right) obtained with two different chains initialised at overdispersed starting values. Bottom panel: mean voxel-wise posterior activation center intensity $\sum_{i=1}^{I}{\left[\lambda_i^0(v)+\rho_i(v)\right]}$ at axial slices $z=-22$ (left), $z=-16$ (middle) and $z=-2$ (right), as obtained from the second run of the model.}
      \label{fig:diagnostics}
 \end{sidewaysfigure}

In this Section, we provide some convergence diagnostics for the BHICP analysis of the emotions dataset. 
In particular, we run a second chain from different starting values, and compare the outputs to those of the original run in Section~\ref{sec:reviewreal} to check if results agree. 
In the top panel of Figure \ref{fig:diagnostics} we see draws from the posterior distribution of the total number of population centers (left), the integrated intensity in the left amygdala (middle), and the integrated intensity in the right amygdala (right). 
Traceplots from the two runs appear to be very similar, and all three parameters seem to have reached their stationary distribution. 
In the bottom panel of Figure \ref{fig:diagnostics} we present the activation center intensity at the same axial slices of Figure \ref{fig:emotionresults}. 
We find that the obtained statistics only present very minor differences. 
Overall, we find no evidence against the convergence of our chains based on our diagnostics. 
However, in the analysis of new datasets it is recommended to additionally consider more formal diagnostics (see, for example, \citet{Kang2011}).    

\end{document}